\documentclass[fleqn,usenatbib]{mnras}
\usepackage{newtxtext,newtxmath}
\usepackage[T1]{fontenc}
\usepackage{ae,aecompl}

\usepackage[dvipdfmx]{graphicx}	% Including figure files
\usepackage{amsmath}	% Advanced maths commands
\usepackage{todonotes,ifthen}
\newcounter{mycomment}
\newcommand{\mycomment}[2][]{%
    \refstepcounter{mycomment}%
    \ifthenelse{\equal{#1}{SS}}
    {\todo[color={red!40},size=\small, inline]{% 
     \textbf{{#1}\themycomment:}~#2}%
    }{\ifthenelse{\equal{#1}{SdlT}}
    	{\todo[color={blue!40},size=\small, inline]{% 
     	 \textbf{{#1}\themycomment:}~#2}%
    	}{\ifthenelse{\equal{#1}{OI}}
        {\todo[color={green!40},size=\small, inline]{% 
     	 \textbf{{#1}\themycomment:}~#2}%
    	}{}}}\noindent}

\newcommand{\angstrom}{\text{\normalfont\AA}}

\title[EL-COSMOS catalog]{The Synthetic Emission Line COSMOS catalog: H$\alpha$ and [OII] galaxy luminosity functions and counts at $0.3<z<2.5$}

\author[S.~Saito et al.]{
\parbox{\textwidth}{
Shun Saito,$^{1,2,3}$\thanks{E-mail: saitos@mst.edu}
Sylvain de la Torre,$^{4}$
Olivier Ilbert,$^{4}$
C\'edric Dubois,$^{4}$
Kiyoto Yabe$^{3}$,
and Jean Coupon$^{5}$}
\\
\\
$^{1}$Institute for Multi-messenger Astrophysics and Cosmology, Department of Physics,\\
Missouri University of Science and Technology, 1315 N. Pine Street, Rolla, MO 65409, U.S.A.\\
$^{2}$Max-Planck-Institut f\"{u}r Astrophysik, Karl-Schwarzschild-Star{\ss}e 1, D-85740 Garching bei M\"{u}nchen, Germany\\
$^{3}$Kavli Institute for the Physics and Mathematics of the Universe (WPI), Todai Institutes for Advanced Study,\\
the University of Tokyo, Kashiwanoha, Kashiwa, Chiba 277-8583, Japan\\
$^{4}$Aix Marseille Univ, CNRS, CNES, LAM, Marseille, France\\
$^{5}$Astronomical Observatory of the University of Geneva, ch. 
d'Ecogia 16, CH-1290 Versoix, Switzerland\\
}

\date{Accepted XXX. Received YYY; in original form ZZZ}

\pubyear{2020}

\begin{document}
\label{firstpage}
\pagerange{\pageref{firstpage}--\pageref{lastpage}}
\maketitle

\begin{abstract}
Star-forming galaxies with strong nebular and collisional emission lines are privileged target galaxies in forthcoming cosmological large galaxy redshift surveys. 
We use the COSMOS2015 photometric catalog to model galaxy spectral energy distributions and emission-line fluxes. 
We adopt an empirical but physically-motivated model that uses information from the best-fitting spectral energy distribution of stellar continuum to each galaxy. 
The emission-line flux model is calibrated and validated against direct flux measurements in subsets of galaxies that have 3D-HST or zCOSMOS-Bright spectra. 
We take a particular care in modelling dust attenuation such that our model can explain both H$\alpha$ and [OII] observed fluxes at different redshifts.
We find that a simple solution to this is to introduce a redshift evolution in the dust attenuation fraction parameter, $f=E_{\rm star}(B-V)/E_{\rm gas}(B-V)$, as $f(z)=0.44+0.2z$.
From this catalog, we derive the H$\alpha$ and [OII] luminosity functions up to redshifts of about 2.5 after carefully accounting for emission line flux and redshift errors.
This allows us to make predictions for H$\alpha$ and [OII] galaxy number counts in next-generation cosmological redshift surveys. 
Our modeled emission lines and spectra in the COSMOS2015 catalog shall be useful to study the target selection for planned next-generation galaxy redshift surveys and we make them publicly available as `EL-COSMOS' on the  \href{http://cesam.lam.fr/aspic/}{ASPIC} database.
\end{abstract}


%%%%%%%%%%%%%%%%%%%%%%%%%%%%%%%%%%%%%%%%%%%%%%%%%%

%%%%%%%%%%%%%%%%% BODY OF PAPER %%%%%%%%%%%%%%%%%%

%==============================%
%==============================%
\section{Introduction}
\label{sec:introduction}
%==============================%
%==============================%
In most of modern cosmological surveys, it is key to map out the large-scale structure of the Universe at high redshift through \textit{emission lines (ELs)} from young star-forming galaxies.
%Identifying the origin of cosmic acceleration is one of the central issues in modern cosmology \citep[for a review, see][]{Weinberg:2012uq}. 
%Although the cosmological constant is consistent with low redshift observations at $z\lesssim 1$ \citep[e.g.,][]{Alam:2017qf,DES-Collaboration:2017ij}, it is a key to map out the evolution of both cosmic expansion and growth of the large-scale structure towards higher redshift. 
Particularly, ongoing and planned galaxy redshift surveys such as the extended Baryonic Oscillation Spectroscopic Survey \citep[eBOSS,][]{Dawson:2016aa}, the Hobby-Eberly Telescope Dark Energy Experiment \citep[HETDEX,][]{Adams:2011ly,Hill:2008aa}, the Dark Energy Spectroscopic Instrument \citep[DESI,][]{DESI-Collaboration:2016tg}, the Subaru Prime Focus Spectrograph \citep[PFS,][]{Takada:2014sf}, Wide-Field InfraRed Survey Telescope-Astrophysics Focused Telescope Assets \citep[WFIRST-AFTA,][]{Spergel:2013aa} and Euclid \citep[][]{Laureijs:2011aa}, are all designed to target star-forming galaxies with strong nebular and collisional emission lines at $z\gtrsim 1$.
PFS and DESI will observe [OII] emitters at $0.6\lesssim z \lesssim 2.4$ and at $0.6 \lesssim z \lesssim 1.7$, respectively, while Euclid will target H$\alpha$-emitting galaxies at $0.9 \lesssim z \lesssim 1.8$ as well as [OIII] emitters at $z>1.5$.
In addition, there are growing interests in performing intensity mapping surveys, where the large-scale structure is traced efficiently by observations with low spatial and wavelength resolutions \citep[for a review, see][]{Kovetz:2017xx}. 
One outstanding example is the Spectro-Photometer for the History of the Universe, Epoch of Reionization, and Ices Explorer \citep[SPHEREx, 2021-,][]{Dore:2014ss} which should detect H$\alpha$-, H$\beta$-, [OII]-, and [OIII]-emitters in a wide redshift range. 

To prepare and optimize those surveys, it is essential to predict the number density of emission-line galaxies (ELGs) for the targeted ELs and redshifts.
Since ELGs are usually primarily selected upon galaxy properties derived from broadband imaging data, it is insufficient to only know the luminosity function and mandatory \textit{to model emission line fluxes as a function of galaxy properties} such as color, star formation rate (SFR), stellar mass and dust attenuation. 
For instance, the PFS cosmological survey will select [OII]-emitter candidates solely from the 5-band photometry, ($g,r,i,z,y$), of the Subaru Hyper Suprime-Cam survey \citep[HSC,][]{Miyazaki:2018aa}.
From a galaxy evolution point of view, it is crucial to connect EL fluxes with other galaxy physical properties, because emission lines trace the photo-ionized gas in the interstellar medium (ISM) regions of galaxies \citep[see e.g.,][]{Draine:2011aa,Conroy:2013aa}. 
This is why it is standard to investigate EL fluxes and ratios as a function of wide range of galaxy properties in the observations \citep[e.g.,][]{Shapley:2005bb,Yabe:2012aa, Kashino:2013jk,Kashino:2018aa,Puglisi:2016aa,Wisnioski:2019aa} as well as in numerical simulations 
\citep[e.g.,][]{Hirschmann:2017tt}. 

Furthermore, the connection between EL fluxes and other galaxy properties is useful not only to constrain galaxy formation physics but also for cosmology. 
Predicting the ELG number density is not only useful to optimize future surveys, but also to be able to analyse and interpret their data. 
For instance, the strength of galaxy clustering on large scales, dubbed galaxy bias, is determined mainly by the mass of host dark matter halos, which tightly correlates with stellar mass \citep[e.g.,][]{Coupon:2015aa}. 
Therefore, one can investigate the selection of ELGs in terms of host halo mass and hence its impact on the galaxy bias \citep[for the galaxy-halo connection of a color-selected sample, see e.g.,][]{Saito:2016oj,Guo:2019ee,Alam:2019aa}.\par 

It is however cumbersome to tackle the EL flux modeling from first principles, given the lack of a complete physical understanding of the interstellar medium. 
Instead, it is more practical to derive an empirical model from actual observational data of galaxy samples that contain multiple broadband photometry as well as spectroscopic measurements of emission lines. 
Fortunately, there are suitable datasets publicly available in the COSMOS field. \citet[][hereafter J09]{Jouvel:2009qy} initiated an effort to model the emission-line fluxes by making use of the multi-band photometry in the COSMOS field, which was used to perform forecasts for several future surveys \citep[e.g.,][]{Takada:2014sf,Newman:2015aa,Pullen:2016aa}.

The aim of this paper is to provide a representative sample of galaxies with predicted emission-line fluxes as a function of redshift and other galaxy properties, 
%, assuming that the COSMOS field is a fairly good representation of the Universe
and provide a new reference catalog of modeled galaxy spectral energy distributions (SEDs).
%on the basis of the latest and deepest COSMOS photometric catalog \citep{Laigle:2016fr}.
In this paper we mainly focus on discussing H$\alpha\,\lambda6563$, H$\beta$\,$\lambda4861$, [OII]$\,(\lambda 3726, \lambda 3729)$ and [OIII]$\,(\lambda 4959,\lambda 5006)$ emission lines. 
Unless specifically mentioned, in the following we refer to the [OII] doublet simply as [OII] and to the [OIII] $\lambda$ 5006 line as [OIII]. 
Note that we quote the wavelength of ELs in air and in units of angstrom (\AA) as opposed to \citet{Draine:2011aa}.
This paper provides major improvements with respect to the work of J09. Firstly, we make use of the most updated COSMOS2015 photometric catalog described in \citet{Laigle:2016fr}, as opposed to the previous \citet{Capak:2007gf} catalog used in J09. 
Secondly, we improve the emission line modeling using a more careful treatment. 
In particular, we calibrate and validate our approach against spectroscopic measurements of mission line fluxes, addressing the impact of dust attenuation. 
Finally, we derive emission line luminosity functions (LFs) taking care of the impact of our modeling uncertainties, which we later compare with various previous measurements from the literature.

Let us briefly mention recent studies on a similar topic. 
\citet{Valentino:2017nx} estimated the H$\alpha$, H$\beta$, [OII] and [OIII] line fluxes in a similar manner to our approach in COSMOS and GOODS-S fields. 
Since their calibration is based on the FMOS-COSMOS survey \citep[][]{Silverman:2015aa}, their prediction is limited at a particular redshift of $z\sim ~1.6$ \citep[see also][]{Kashino:2018aa}. 
\citet{Merson:2017bu} predicted H$\alpha$ number counts in a semi-analytic galaxy model of \texttt{GALACTICUS} by running the photo-ionization code of \texttt{CLOUDY}. 
Similarly, \citet{Izquierdo-Villalba:2019aa} computed the EL fluxes in the semi-analytic \texttt{Millenium} simulation, following a recipe in \citet{Orsi:2014aa} where the HII region in a galaxy is modeled as a function of metallicity. 
We note however that \citet{Izquierdo-Villalba:2019aa} calibrated the dust attenuation part for ELs such that it reproduces the observed luminosity functions. 
In contrast, the luminosity functions are purely predictions in our case. 
The main advantage of our approach is that our synthetic EL catalog spans uniquely a wide redshift range of $0\lesssim z \lesssim 2.5$, and that it naturally provides an empirical relation between global galaxy properties and EL fluxes of interest.
Also, it is worth noting that those previous works use generic dust attenuation laws, although there are slight differences in the detail. 
We will show how significantly the dust attenuation calibration matters and address the differences between our approach and these relevant works in the main text, when appropriate.

The paper is structured as follows. 
We introduce the observational dataset that we adopt in section~\ref{sec:observation}. 
In section~\ref{sec:model}, we first explain the modeling of galaxy spectra, including both stellar continuum and ELs. After discussing the calibration required to match to observed EL fluxes, we show the performance of our modeling and discuss the potential limitations in our approach. In section~\ref{sec:lf}, we describe the uncertainties in the modeled EL luminosities and present the predicted H$\alpha$ and [OII] luminosity functions. The number counts for planned galaxy surveys are discussed in section~\ref{sec:discussion}.  
Finally, we conclude in section~\ref{sec:conclusion}. 
Throughout this paper, we refer to our synthetic emission-line predictions for the COSMOS galaxies as the EL-COSMOS catalog. 
Unless otherwise specified, we assume a flat $\Lambda$CDM cosmology with $\Omega_{\rm m}=0.3$ and $H_{0}=70\,{\rm km\,s^{-1}\,Mpc^{-1}}$.

%~~~~~~~~~~~~~~~~~~~~~~~~~~~~~~%
%\begin{figure}
%\begin{center}
%\hspace*{-0.7cm}
%\includegraphics[width=1\columnwidth]{smf.eps}
%\caption{
%	\label{fig:smf}
%    The SMF evolution in two different photometry catalogs in the %COSMOS field. J09-CMC was based on \citet{Capak:2007gf}, and its SMF %(\textit{dashed}) evolution is unphysical at a small mass end, %$\log_{10}(M_{*}/M_{\odot})\lesssim 10$. This behavior is not seen in %SMFs from the updated COSMOS2015 catalog in \citet{Laigle:2016fr} %(\textit{solid}). 
%}
%\end{center}
%\end{figure}
%~~~~~~~~~~~~~~~~~~~~~~~~~~~~~~%

%==============================%
%==============================%
\section{Observational data}
\label{sec:observation}
%==============================%
%==============================%

In this section we briefly describe the observational data that we use in this analysis. We first use the COSMOS2015 multi-band photometric catalog of \citet{Laigle:2016fr} for which we model galaxy emission line fluxes. The emission line fluxes are uniquely computed from derived quantities from the COSMOS2015 catalog. In addition, we make use of spectroscopic measurements of emission line fluxes for a subset of galaxies. This enables us to verify and calibrate our predictions. We note that we describe the observations involved in the luminosity function measurements later in Sec.~\ref{sec:lf}.\par

%==============================%
\subsection{COSMOS2015 multi-band photometry}
\label{subsec:photo}
%==============================%

\citet{Laigle:2016fr} provide the most updated photometric catalog in the COSMOS field. 
$31$ photometric bands nearly uniformly cover a wide range of wavelength from near ultraviolet (2000\AA) to mid infrared ($10^{5}$\AA). 
The COSMOS2015 catalog includes a homogeneous population of galaxies essentially selected in near infrared ($K_{s}\le 24.7$ in ultra-deep field).\par 

In this paper, we select 518404 objects over 1.38 ${\rm deg^{2}}$ (corresponding to $\mathcal{A}^{\rm UVISTA}\, \&\,  \mathcal{A}^{\rm !OPT}\, \&\, \mathcal{A}^{\rm COSMOS}$ in Table~7 of \citet{Laigle:2016fr}) that are classified as `galaxies' in the catalog.
Note that, even though X-ray sources are removed in this classification, active galactic nuclei (AGNs) can be still included.
We refer the reader to \citet{Laigle:2016fr} for further detail. 
As we will explain later in Sec.~\ref{sec:model}, we do not use derived quantities such as the stellar mass or the star formation rate from the publicly available catalog. 

%==============================%
\subsection{Spectroscopic measurement of emission line fluxes}
\label{subsec:spec}
%==============================%

\begin{figure}
\begin{center}
\hspace*{-0.7cm}
\includegraphics[width=1\columnwidth]{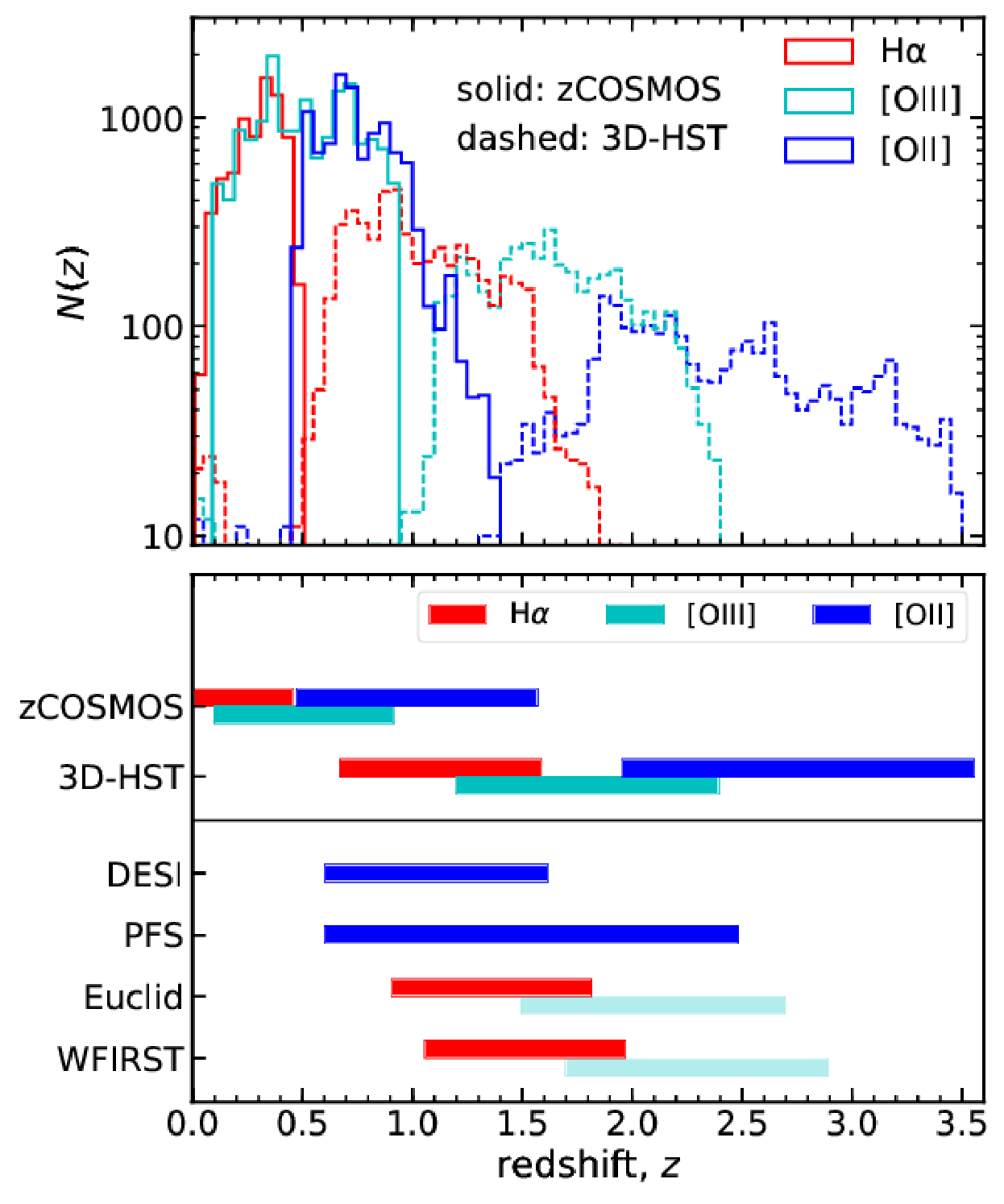}
\hspace*{-0.7cm}
\caption{
	\label{fig:zred_coverage}
    The redshift range covered by the emission lines in each spectroscopic survey. (Top) The number of COSMOS2015 galaxies for which the measurement of each EL line is available as a function of redshift with $\Delta z=0.05$. The solid and dashed histograms correspond to zCOSMOS and 3D-HST, respectively. 
    %The difference in the overall amplitude originates mainly from the difference in their sky coverages (roughly speaking, $\sim 1.3\,{\rm deg^{2}}$ for zCOSMOS, and $\sim 0.03\,{\rm deg^{2}}$ for 3D-HST). \mycomment[OI]{Is this statement really necessary ? Depth and spectroscopic coverage are also really important. I would remove this sentence, or add other factors.} 
    (Bottom) The upper half (3D-HST and zCOSMOS) shows the range in which we calibrate our prediction of the emission-line flux, while the lower half shows the ranges targeted by ongoing and forthcoming galaxy redshift surveys. 
    Since [OIII] is not a main target EL of Euclid and WFIRST, we make it transparent. 
}
\end{center}
\end{figure}

In order to calibrate our method, we make use of two dataset of emission line fluxes from the zCOSMOS-Bright \citep[][]{Lilly:2007cs} and the 3D-HST \citep[][]{Momcheva:2016lr,brammer12} surveys. 
The wavelength coverage of zCOSMOS and 3D-HST is $5500 < \lambda\,[\angstrom] < 9600$ and $11000 < \lambda\,[\angstrom] < 17000$, respectively. 
These correspond to $z \lesssim 0.46$ for H$\alpha$ and $0.47 \lesssim z \lesssim 1.57$ for [OII] in zCOSMOS, and to $0.67 \lesssim z \lesssim 1.59$ for H$\alpha$ and $1.95 \lesssim z \lesssim 3.56$ for [OII] in 3D-HST, respectively. 
This allows us to cover a relatively wide range of redshifts for the strong emission lines [OII] and H$\alpha$ that are targeted in ongoing and future galaxy redshift surveys, as shown in Figure~\ref{fig:zred_coverage}.
However, note that, due to a poor spectral resolution in these surveys, some broadened lines such as the [OII] doublet or H$\alpha$+[NII] complex are not well resolved. 
Nonetheless, the EL fitting procedure took this into account \citep[e.g.][]{Momcheva:2016lr} and quoted EL fluxes are assumed to be free from contamination.
In appendix~\ref{appendix:z_photo}, we show a comparison between photometric and spectroscopic redshifts. 

In the case of 3D-HST, we take the publicly available catalog of emission line measurements in the COSMOS field\footnote{https://3dhst.research.yale.edu/Data.php}. 
We refer to Sec.~6 in \citet[][]{Momcheva:2016lr} for the details of their EL measurements.
We note that $z_{\rm best}$ in the 3D-HST catalog does not always correspond to the redshift spectroscopically confirmed from emission lines \citep[][]{Momcheva:2016lr}.
This is because $z_{\rm best}$ is chosen as the best estimate from photometric and spectroscopic redshifts, and the photometric redshift can be more precise in their measurement. 
As a consequence, the $z_{\rm best}$ distribution in the original catalog is wider than the redshift ranges that ELs are measurable. We do not adopt objects out of the measurable ranges, and make a sharp cut in redshift, as shown in Figure~\ref{fig:zred_coverage}. 
We then match the objects with those in the COSMOS2015 photometry catalog within an angular radius of $0.3$ arcsec. 
%Since our predictions are based on photometric redshifts, we make a further cut in term of photometric redshift even though a spectroscopic redshift is available. \mycomment[OI]{Could you specify the last sentence. That's unclear to me.}

We also use the zCOSMOS catalog of emission line measurements presented in \citet[][]{Silverman:2009} but extended to the final zCOSMOS-Bright spectroscopic sample containing about 20,000 galaxies with reliable redshifts. 
The EL measurement is performed by the automated pipeline \texttt{PLATEFIT} whose detail is described in \citet[][]{Lamareille:2009tt,Argence:2009aa}.
We select the objects in the zCOSMOS catalog that match those in COSMOS2015 and have the most reliable redshifts and flux measurements, i.e., with $\texttt{zflag}\in [3.1,3.4,3.5,4.1,4.3,4.4,4.5,9.3,9.4,9.5]$ following \citet[][]{Lilly:2007cs}.
Also, we need to take into account the fact that there is a loss in the measured EL fluxes due to the finite aperture of a slit or a fiber.
Such an aperture correction for each emission line flux has been applied on an object-by-object basis following the procedure of \citet[][]{Lamareille:2009aa}.
There are two measurements of the aperture correction factors: Subaru and ACS. 
Since the majority of the matched objects possess the corrections from Subaru, we adopt the Subaru ones when available, the ACS ones for the others, and discard objects if none of the two is available.

\begin{figure*}
\begin{center}
\hspace*{-0.7cm}
\includegraphics[width=1\columnwidth]{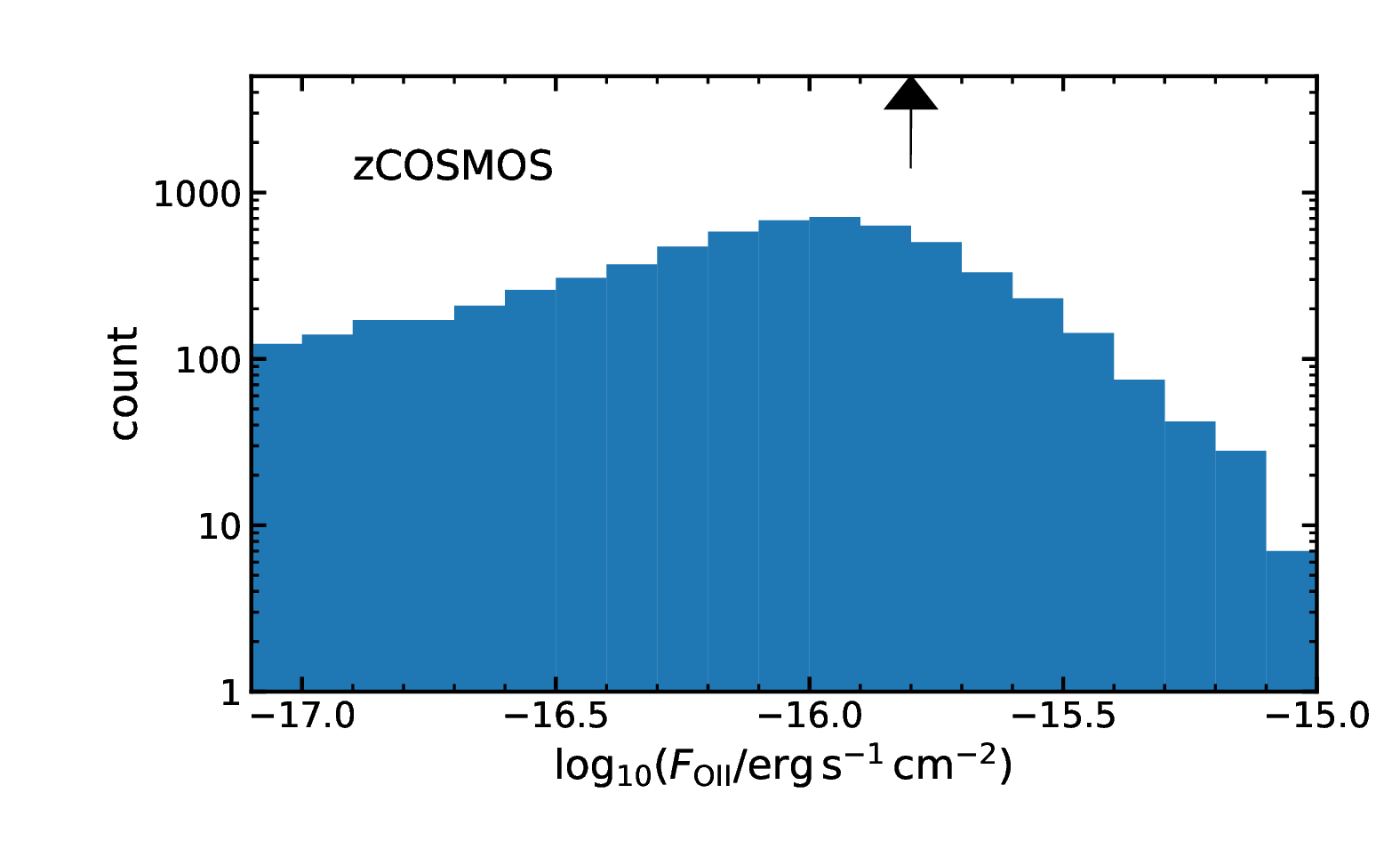}
\includegraphics[width=1\columnwidth]{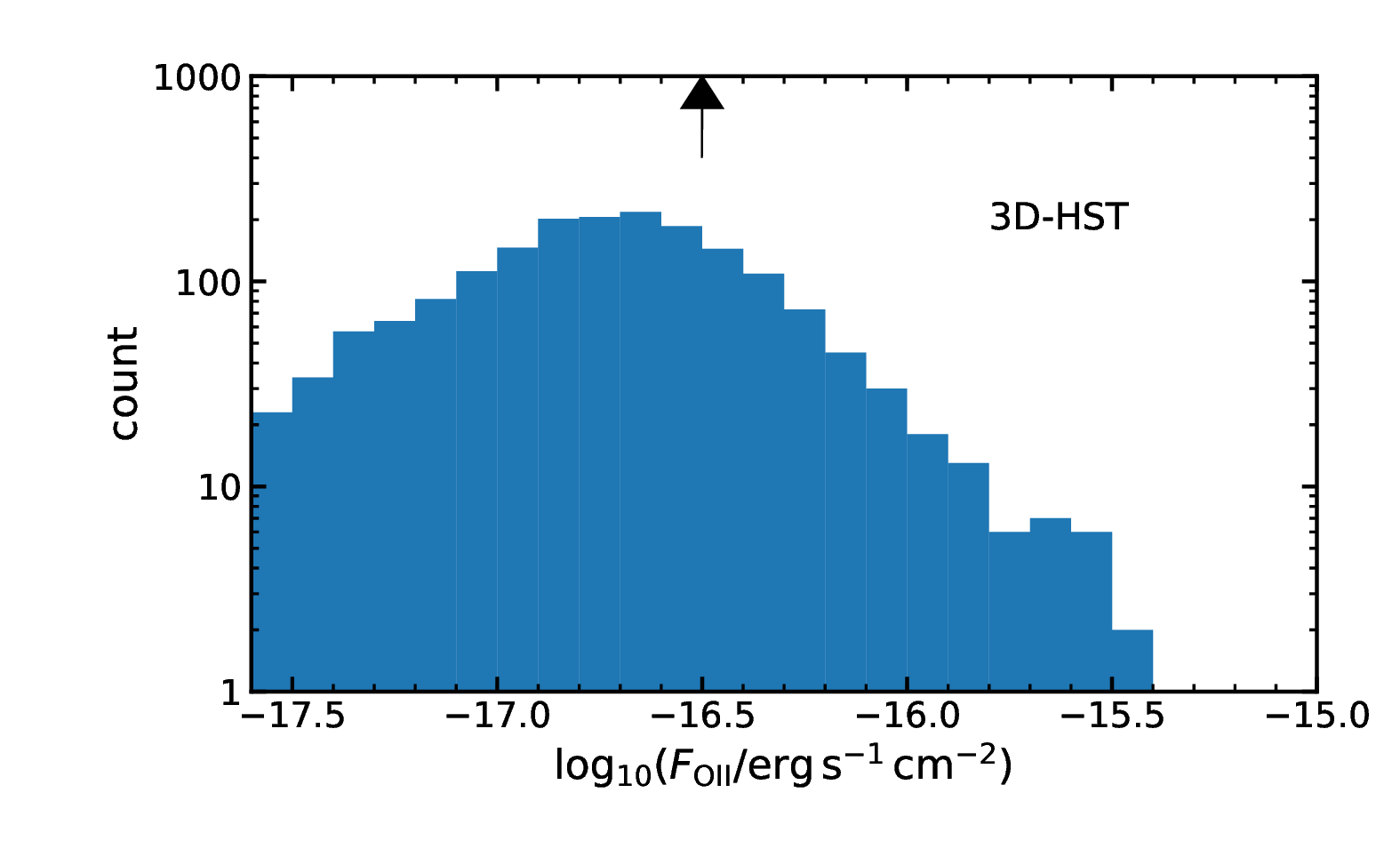}
\caption{
  \label{fig:completeness} 
  The flux incompleteness of zCOSMOS (\textit{left}) and 3D-HST (\textit{right}).
  We quote the flux completeness simply from the histogram of the [OII] flux measurements: $\log_{10}(F/{\rm erg\,s^{-1}\,cm^{-2}})=-15.8$ and -16.5 for zCOSMOS and 3D-HST, respectively, as indicated by arrows. 
}
\end{center}
\end{figure*}

Since the COSMOS2015 photometric catalog contains galaxies whose emission line fluxes are smaller than the completeness limit in zCOSMOS and 3D-HST, we can study the selection bias in our prediction. 
We define an approximate completeness limit in each flux measurement. 
Figure~\ref{fig:completeness} shows the histogram of the [OII] flux measurement from the two redshift surveys, showing a clear declination towards lower flux due to incompleteness. 
In Section~\ref{sec:model}, we consider all the galaxies in the catalogs without imposing the further cut with e.g., the signal-to-noise of the EL measurements, and hence include galaxies down to very faint fluxes.
Meanwhile, we limit ourselves to the objects with emission lines of $S/N>2.5$ for the luminosity function analysis in Section~\ref{sec:lf}, since the luminosity function is sensitive to errors as we address in detail. 
A strict value of the completeness limit should be evaluated with a reference complete sample which is not always available. 
Nevertheless we simply define here a completeness limit from the histogram in Figure~\ref{fig:completeness} as the point where counts start to deviate from power law by a factor of more than 50\%.
As shown by arrows in Figure~\ref{fig:completeness}, this definition gives the completeness limit as $\log_{10}(F/{\rm erg\,s^{-1}\,cm^{-2}})=-15.8$ and -16.5 for zCOSMOS and 3D-HST, respectively. 
Note that, in the case of 3D-HST, our value is consistent with the $\sim 80$\% completeness limit from the H$\beta$ measurement in \citet{Zeimann:2014aa} (see their Figure~2).

%~~~~~~~~~~~~~~~~~~~~~~~~~~~~~~%
%~~~~~~~~~~~~~~~~~~~~~~~~~~~~~~%

%~~~~~~~~~~~~~~~~~~~~~~~~~~~~~~%
%~~~~~~~~~~~~~~~~~~~~~~~~~~~~~~%

%==============================%
%==============================%
\section{Modeling the emission line flux}
\label{sec:model}
%==============================%
%==============================%

%==============================%
\subsection{Overview}
\label{subsec:overview}
%==============================%

We describe in this section the way we model the galaxy spectral energy distribution including ELs. 
We emphasize that the difference from \citet{Laigle:2016fr} is only the modeling of ELs, and that the fitting procedure follows the same way as \citet{Laigle:2016fr} performed; the template is computed with the \texttt{LEPHARE} code, and the bestfit template is chosen as the one with the minimum $\chi^{2}$, compared with the multi-band photometric data (see Table~1 in \citet[][]{Laigle:2016fr}). 
All the fitting was performed on a flux-based approach (rather than a magnitude-based) for which non-detected photometric data is straightforwardly dealt with \citep[see Sec.~3 of][]{Laigle:2016fr}.

First of all, we model the stellar continuum spectrum in exactly the same manner as that in \citet{Ilbert:2015aa,Laigle:2016fr}. 
We adopt templates from a stellar population synthesis (SPS) model that includes a wide variety of synthetic galaxy spectra in \citet[][hereafter BC03]{Bruzual:2003aa}, assuming a \citet{chabrier03} initial mass function. 
We further convolve the SPS template with two kinds of star formation histories (SFHs), either an exponentially declining SFH, $e^{-t/\tau}$, or a delayed SFH, $(t/\tau)^{2}e^{-t/\tau}$, where we consider as free parameters the galaxy age, $t$, within a range of 0.5-13.5 Gyr, and SFH timescale, $\tau$, within a range of 0.1-30 Gyr \citep[][]{Ilbert:2010xp}.
In terms of dust reddening, we apply the attenuation factor, $10^{-0.4k(\lambda)E_{\rm star}(B-V)}$ with two possible forms for $k(\lambda)$, either a curve with slope of $\lambda^{-0.9}$ as in \citet[][]{Arnouts:2013aa}, or the starburst curve of \citet{Calzetti:2000kx}. 
The color excess for the stellar continuum, $E_{\rm star}(B-V)$, is a free parameter and is allowed to take discrete values in the range of 0-0.7 with a step size of 0.1. We also consider two metallicities, $Z_{\odot}$ and $0.5Z_{\odot}$. 
In summary, we have 12 (BC03 templates) $\times 43$ (ages) $\times 12$ (SFH and metallicities) $\times$ $2\times 8$ (dust attenuation) $=99,072$ templates in modeling the stellar continuum.\par

In addition, we add to the BC03 spectra the contribution of the star-forming nebular regions in term of the continuum emission and discrete ELs \citep[][]{Schaerer:2009lr}. 
We first compute the number of Lyman continuum photons, $Q_{\rm LyC}$, by integrating the luminosity of the BC03 spectra up to a given wavelength and dividing it by a corresponding energy (e.g., 13.6eV, 24.59eV and 54.42eV for HI, HeI, and HeII, respectively). 
Following \citet{Schaerer:1998aa}, we then convert it to the monochromatic luminosity of the nebular gas as
%%%%%%
\begin{equation}
 L_{\lambda} = \frac{hc}{\lambda}\frac{\alpha_{\lambda}(T_{e})}{\alpha_{B}(T_{e})}\; f_\gamma \; Q_{\rm LyC},   
\end{equation}
%%%%%%
where, for simplicity, we assume that the emitting gas has an electron temperature of $T_{e}=10^{4}\,{\rm K}$ and an electron density of $n_{e}=100\,{\rm cm^{-3}}$ where the case B recombination coefficient for hydrogen is given by $\alpha_{B}=2.59\times 10^{-13}\,{\rm cm^{3}s^{-1}}$. 
$f_\gamma$ denotes the fraction of ionizing photons absorbed by the gas. We assume all ionizing photons are absorbed, i.e., $f_\gamma=1$. 
To compute the nebular continuum, we compute the continuum coefficient, $\alpha_{\lambda}$, by further assuming that $n(\rm HeII)/n(\rm HI)=0.1$ and $n(\rm HeIII)/n(\rm HI)=0$ and accounting for free-free and free-bound emission by hydrogen and neutral helium as well as the two-photon continuum of hydrogen \citep[][]{Krueger:1995aa}.
To compute the discrete ELs, we specifically derive the ${\rm H\beta}$ luminosity \citep[][]{Krueger:1995aa,Osterbrock:2006aa}:
%%%%%%
\begin{equation}
L^{\rm int}_{\rm H_\beta} = 4.78 \times 10^{-13} \; f_\gamma \; Q_{\rm LyC}.
\end{equation}
%%%%%%
We compute luminosity for other lines by simply assuming the line ratios for a metallicity of $Z > 0.2Z_{\odot}$ from Table~1 in \citet{Anders:2003aa} for the non-hydrogen lines and Table~4.2 in \citet{Osterbrock:2006aa} for the hydrogen ones. 
Note that, since the two metallicities, $Z_{\odot}$ and $0.5Z_{\odot}$ lie within the range of $Z > 0.2Z_{\odot}$, the line ratios are essentially fixed with ${\rm [OII]}/{\rm H\beta} = 3$, ${\rm H\alpha}/{\rm H\beta} = 2.9$ and ${\rm [OIII]}/{\rm H\beta} = 4.1$.
As one exceptional case, we also attempt to make the [OIII] luminosity \textit{free} by further multiplying a free scaling parameter as ${\rm [OIII]}/(4.1{\rm H\beta}) = 0.25$, $0.5$, $1$, $2$, and $4$. 
This is motivated by the fact that sufficient information is present in the photometry to provide a coarse constraint on the emission line flux, in the wavelength regime covered by medium-band filters. Therefore we will present separately the [OIII] results in section~\ref{subsubsec:flux_o3}. 
We then add the ELs to the continuum assuming a Gaussian form with a rotational velocity of $V_{\rm rot}=200 {\rm km/s}$, and with a fixed inclination angle of $<\sin \; i>=0.5$.\par 

We account for the fact that the amount of dust attenuation for the ELs from nebular regions can be different from one for the stellar continuum.
Following \citet[][]{Calzetti:1994aa}, we adopt the extinction curve $k(\lambda)$ for the Milky Way from \citet[][]{Seaton:1979aa} for the ELs, but parametrize the amplitude of color excess by a factor of $f$ such that:
%%%%%%
\begin{equation}
 E_{\rm neb}(B-V) = \frac{E_{\rm star}(B-V)}{f}. 
 \label{eq:EBVneb}
\end{equation}
%%%%%%
In general, the measurements of the Balmer decrement (i.e., ${\rm H\alpha/H\beta}$) and comparison between SFRs from H$\alpha$ and UV have shown that the amount of dust attenuation for nebula ELs are generally larger than one for the stellar component, i.e., the factor, $f$, is smaller than one at a low redshift. 
Also, $f$ tends to be larger at higher redshift. 
An excellent summary of recent such measurements in the literature is presented in Table 3 in \citet[][]{Puglisi:2016aa}: 
\citet[][]{Calzetti:1994aa} initiated such an effort, finding that $f(z=0)=0.44\pm 0.03$ \citep[see also][]{Calzetti:2000kx}. 
Also, \citet[][]{Kashino:2013jk} and \citet[][]{Price:2014aa} reported 0.69-0.83 and $0.55 \pm 0.16$ at $z\sim 1.4$, respectively. 
Even at higher redshift, $1.5<z<2.6$, \citet[][]{Reddy:2010aa} found that $f$ is consistent with 1 for their sample of Lyman Break Galaxies.  
\citet[][]{McLure:2018aa} adopted $f=0.76$ to extrapolate the relation between UV attenuation and stellar mass for the $z\sim 1.4$ galaxies in \citet[][]{Kashino:2013jk}, finding that both the $z\sim 1.4$ galaxies and $z\sim 2.5$ galaxies in their ALMA observation seem to follow the dust attenuation law in \citet[][]{Calzetti:1994aa}. 
\citet[][]{Faisst:2019aa} tested the trend of $f$ closer to one at higher redshift by examining local galaxies with a strong H$\alpha$ emission analogous to high-z galaxies. 
Thus, detailed dependence of the dust attenuation in nebular regions on galaxy properties is still open to debate.  
Nevertheless, we will include the redshift dependence, $f(z)$, in an empirical manner as we will explain in section~\ref{subsec:calibration}. 

Finally, the total galaxy spectrum is redshifted and dimmed by the luminosity distance where we adopt a photometric redshift in \citet{Laigle:2016fr}.
Namely, we did not treat the photometric redshift as a free parameter in the SED fitting.  

% %==============================%
% \subsection{The specific case of [OIII]}
% \label{subsec:OIII}
% %==============================%

% \mycomment[OI]{Not sure yet that we need a specific section, but I write what we did anyway. text done quickly. to be modified.}

% We first predict the [OIII] flux as described in \ref{subsec:overview}. However, the comparison between the [OIII] predicted and directly measured in spectra showed averaged differences by more than a factor X. Therefore, we adopted a specific treatment for the [OIII] EL. 

% We allow the [OIII] EL flux to be rescaled by a free factor among 0.25, 0.5, 1, 2, 4. Therefore, predicted magnitudes obtained with BC03 templates include these 5 different configurations for the [OIII] line fluxes. We take as [OIII] EL flux the one producing the better fit of the photometric data.

% As we will show in the next section, such procedure improve significantly the quality of our results, specially in the redshift range where [OIII] is falling within the medium band coverage.

%~~~~~~~~~~~~~~~~~~~~~~~~~~~~~~%
\begin{figure}
\begin{center}
\hspace*{-0.5cm}
\includegraphics[width=0.9\columnwidth]{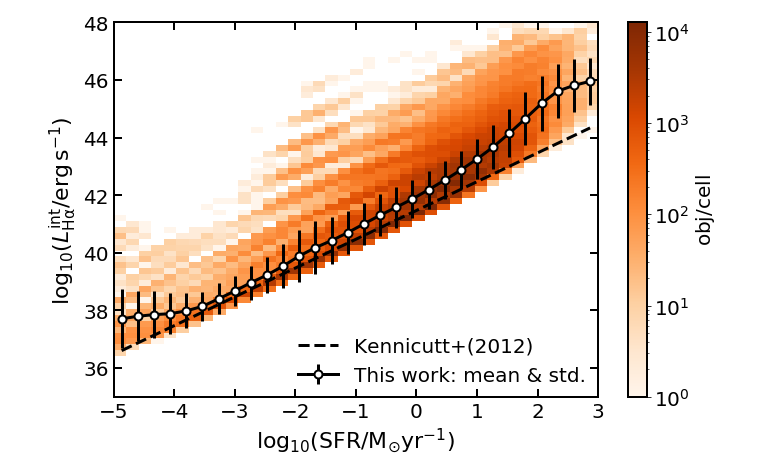}
%\hspace*{-0.7cm}
\caption{
  \label{fig:kennicutt}
  The intrinsic H$\alpha$ luminosity in our fiducial model (presented in section~\ref{subsec:results}) in light of the Kennicutt calibration \citep[dashed line,][]{Kennicutt:2012aa}. 
  We show the mean and standard deviation of our prediction in COSMOS2015 (black points with error bars). For reference, galaxy number densities in each cell of SFR and H$\alpha$ luminosity is color coded as indicated by the color bar. Our intrinsic H$\alpha$ luminosity prediction is in an excellent agreement with \citep[][]{Kennicutt:2012aa} calibration. 
}
\end{center}
\end{figure}
%~~~~~~~~~~~~~~~~~~~~~~~~~~~~~~%

%~~~~~~~~~~~~~~~~~~~~~~~~~~~~~~%
\begin{figure*}
\begin{center}
\hspace*{-0.7cm}
\includegraphics[width=1.1\textwidth]{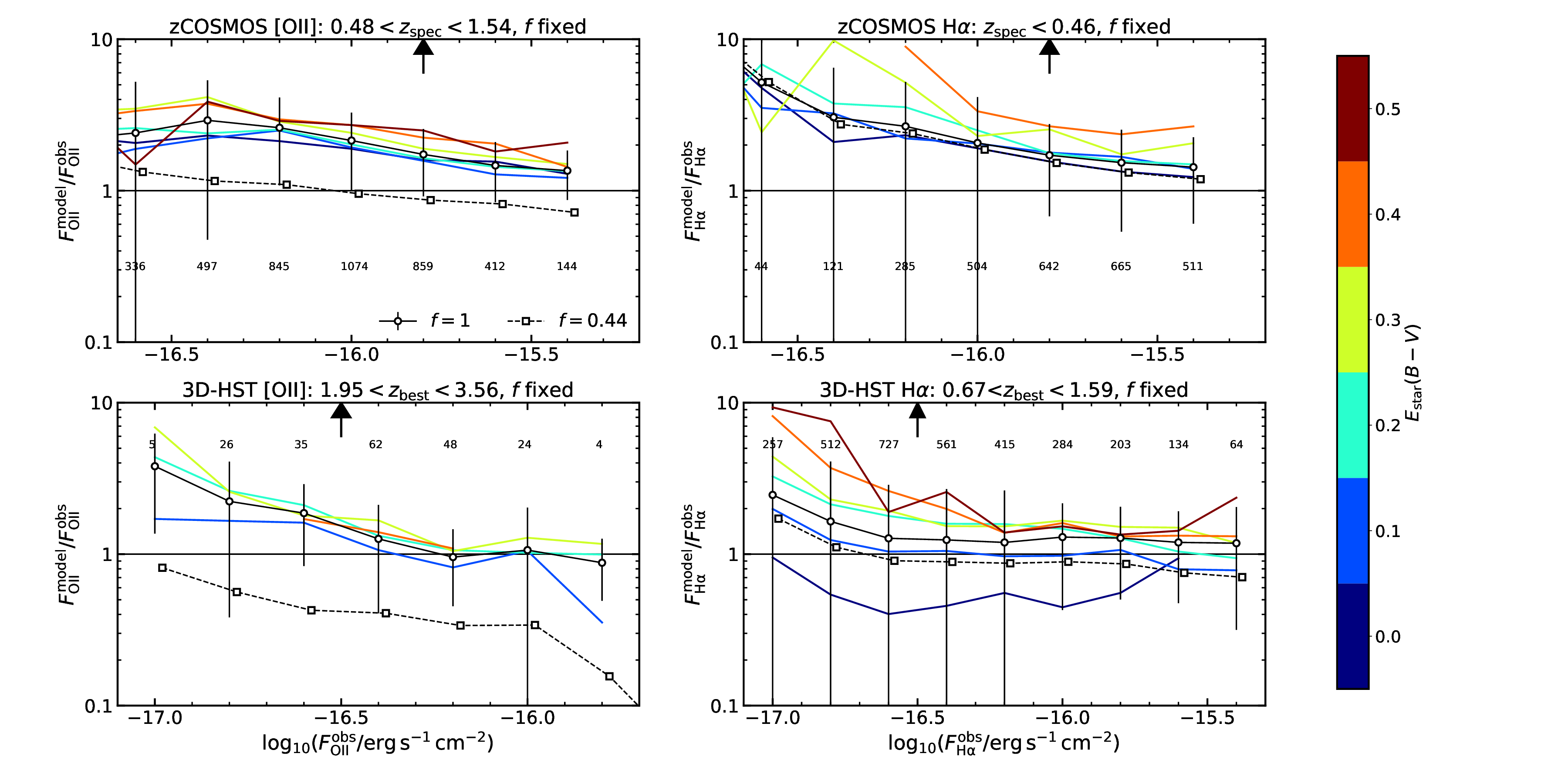}
\caption{
  \label{fig:EBV}
  The performance of the simplest model when we fix the value of dust attenuation factor, $f=1$ (circles and solid lines). 
  The upper panels show the zCOSMOS results for [OII] (left) and H$\alpha$ (right), while the bottom panels show the 3D-HST ones in the same manner. 
  The results with all galaxies for a given sample are shown with the black circles with error bars, while the ones for the subsample binned with the color excess, $E_{\rm star}(B-V)$, are shown with different colors. 
  Note that all the plots are made with a binning size of $\Delta\log_{10}F^{\rm obs}=0.2$ for a fair comparison, and the sample size in each bin is explicitly annotated. 
  We also show the case with $f=0.44$ only for the mean (square data points with dashed lines). 
  The arrows indicate the rough flux limits in Figure~.\ref{fig:completeness}.
}
\end{center}
\end{figure*}
%~~~~~~~~~~~~~~~~~~~~~~~~~~~~~~%

%~~~~~~~~~~~~~~~~~~~~~~~~~~~~~~%
\begin{figure}
\begin{center}
\hspace*{-0.7cm}
\includegraphics[width=1.\columnwidth]{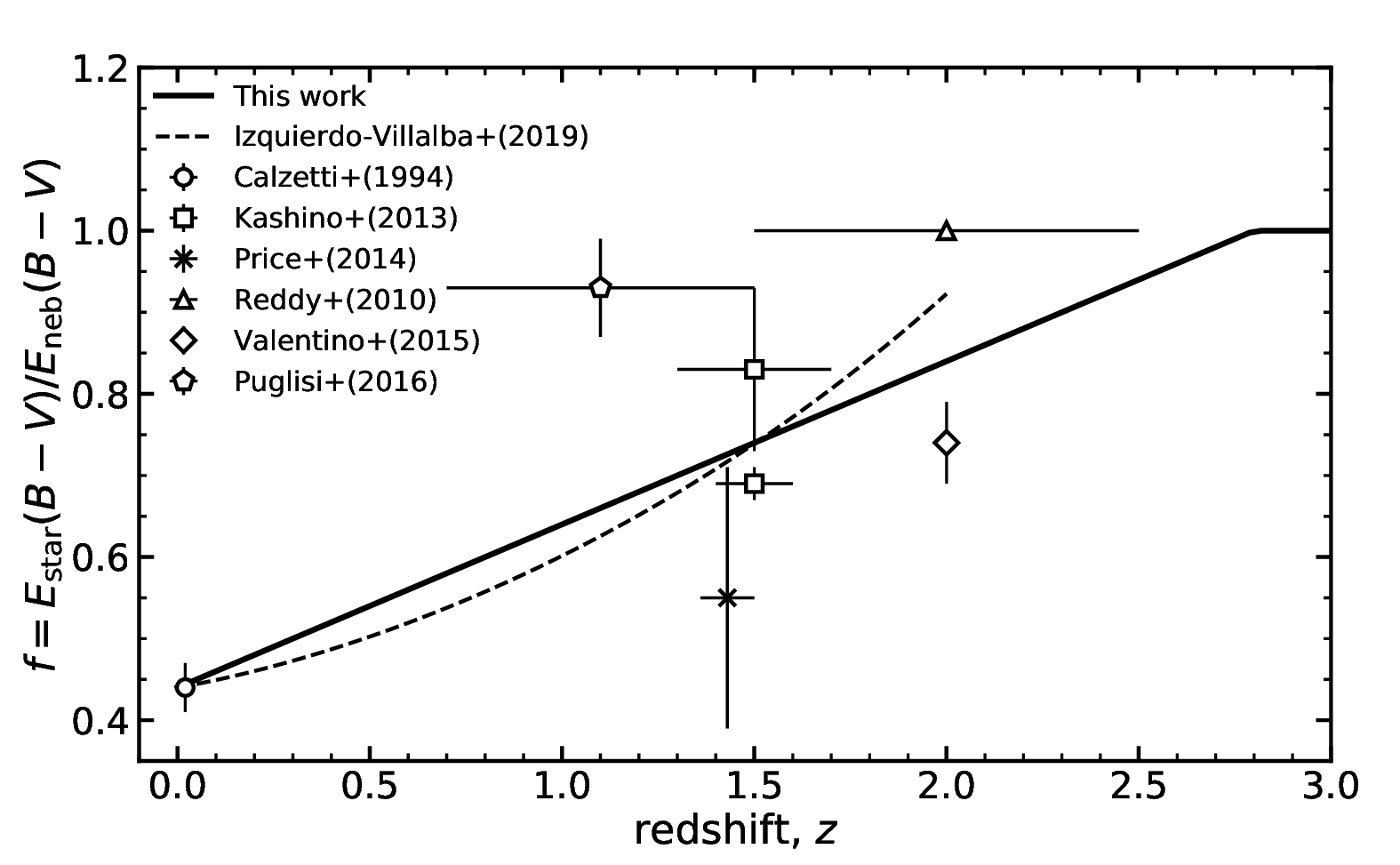}
\hspace*{-0.7cm}
\caption{
  \label{fig:f_z}
  The dust attenuation factor, $f$, as a function of redshift. 
  Our simple proposal, $f(z)=0.44+0.2z$, comes from the spectroscopic EL flux measurements, as discussed later (\textit{solid}). 
  Another curve (\textit{dashed}) is taken from \citet[][]{Izquierdo-Villalba:2019aa}, where the authors calibrated the redshift dependence from the luminosity functions. 
  Note that we normalize their curve such that $f(z=0)=0.44$ to focus on the redshift evolution. 
  The observational data points are mostly taken from \citet[][]{Puglisi:2016aa}. 
  We plot the error bars along $y$ axis when they are available, and the ranges along $x$ axis refer to redshift ranges for each observation. 
}
\end{center}
\end{figure}
%~~~~~~~~~~~~~~~~~~~~~~~~~~~~~~%

%==============================%
\subsection{Calibration: intrinsic EL luminosity and dust attenuation}
\label{subsec:calibration}
%==============================%

It is necessary to calibrate our model to match the directly observed EL fluxes. In particular we need to examine the two main components of our empirical modeling: the intrinsic EL luminosity and the dust attenuation.\par

It is not easy to verify whether our modeling of the intrinsic EL luminosity works well, simply because the intrinsic ELs are not directly observable.
Here we take a slightly different route to examine the plausibility of our modeling.
Among the ELs, H$\alpha$ is regarded as a reliable tracer of the SFR, and the tight correlation between the SFR and the intrinsic (i.e., dust corrected) H$\alpha$ luminosity is well documented in the literature \citep[see e.g.,][]{Kennicutt:1998ab,Kennicutt:1998aa}.
In Figure~\ref{fig:kennicutt}, we compare the relation between our prediction of the intrinsic H$\alpha$ luminosity and the derived SFR with the calibrated relation in the local galaxies \citep{Kennicutt:2012aa}:
%%%%%%
\begin{equation}
 \log_{10}(L^{\rm int}_{\rm H\alpha}/{\rm erg\,s^{-1}}) 
 %= 41.72 - \log_{10}(0.63) + \log_{10}({\rm SFR}/M_{\odot}\,{\rm yr}^{-1}).
 = 41.47 + \log_{10}({\rm SFR}/M_{\odot}\,{\rm yr}^{-1}). 
\end{equation}
%%%%%%
Note that we include a factor of $\log_{10}(0.63)$ to account for the fact that we use the Chabrier IMF rather than the Salpeter one \citep{Madau:2014aa}. 
Here we adopt the fitting result with our fiducial model (described in section~\ref{subsec:results}), but we have checked that the mean relation is not drastically affected by the different model choice. 
The SFR in our COSMOS2015 galaxies is directly encoded in the best-fit template. 
It is clear in Figure~\ref{fig:kennicutt} that our mean prediction of the H$\alpha$ intrinsic luminosity is fully consistent with the Kennicutt calibration within its 1$\sigma$ error.\par

The Kennicutt calibration has been often adopted in the related works;
J09 predicted the intrinsic [OII] luminosity by converting the SFR with the 
relation in \citet{Kennicutt:1998aa}, and computed other ELs with fixed line ratios. 
We caution that \citet{Kennicutt:1998aa} corrected the dust attenuation only at the wavelength of H$\alpha$ and hence their calibration tends to overestimate the intrinsic [OII] luminosity. 
Also, \citet{Valentino:2017nx} adopted the relation between H$\alpha$ and the SFR in \citet{Kennicutt:1998aa}, and computed other ELs with fixed line ratios, e.g., [OII]/H$\alpha=1$. 
Similarly, \citet{Kashino:2018aa} predicted the H$\alpha$ flux with the calibration in \citet{Kennicutt:1998aa} and compared their prediction with the H$\alpha$ measurement in their FMOS sample at $z\sim 1.5$.

As described previously, the intrinsic EL luminosity relies on the number of ionizing photons estimated from the SED template-fitting. 
We show in Figure~\ref{fig:kennicutt} that our modeling of the intrinsic EL luminosity captures well the overall trend in the Kennicutt calibration between H$\alpha$ and SFR. Still, using directly the Kennicutt relation to implement the EL would not produce exactly the same EL, with difference reaching a factor of 0.3dex on average at high SFR. 
Moreover, our model naturally produces a scatter in the EL at a fixed SFR, which includes contributions from both intrinsic scatter and modeling uncertainty due to the SED fitting. 

%There are obvious advantages in our modeling. 
%First, our model of the intrinsic EL luminosity does not rely directly on the SFR, but on the number of ionizing photons.
%The SED template-fitting does not necessarily provide an consistent value with observed SFR \citep{Laigle:2016fr,Laigle:2019aa} 
%Secondly, since we do not force the mean relation in the Kennicutt calibration, our model naturally includes scatter at a fixed SFR. 
%Nonetheless, we stress that the scatter in Figure~\ref{fig:kennicutt} includes contributions from both  intrinsic scatter and modeling error due to the SED fitting to COSMOS2015. 
%We thus argue that our modeling of the intrinsic EL luminosity captures well the overall trend in the Kennicutt calibration. 
%although it is possible to solve the detailed ISM structure in a physically well-motivated manner \citep[see e.g., ][]{Merson:2017bu}. 
%\mycomment[OI]{I would simply rewrite the previous paragraph like this: As described previously, the intrinsic EL luminosity relies on the number of ionizing photons estimated from template-fitting. We show in Figure~\ref{fig:kennicutt} that our modeling of the intrinsic EL luminosity captures well the overall trend in the Kennicutt calibration between $H_\alpha$ and SFR. Still, using directly the Kennicutt relation to implement the EL would not produce exactly the same EL, with difference reaching a factor XX on average at high SFR. Moreover, our model naturally produces scatter in the EL at a fixed SFR, which is useful for our analysis.}

Next let us examine the impact of the dust attenuation with the simplest model with $f=1$, i.e., $E_{\rm neb}(B-V) = E_{\rm star}(B-V)$. 
Figure~\ref{fig:EBV} shows the comparison between our model prediction with $f=1$ and the observed flux. 
We plot the mean and the standard deviation from available samples with black open circles with error bars. 
Note that the standard deviation here includes a statistical uncertainty associated with the sample size (indicated with small numbers in each bin) as well as an intrinsic scatter which encodes the uncertainty in the SED fitting.  
Also, in the case of the high-$z$ [OII] samples in 3D-HST, we restrict ourselves to galaxies whose redshifts are obtained from 3D-HST spectra (\texttt{z\_best\_s==2}) and relatively close to photometric ones in COSMOS2015 as $|z_{\rm spec}-z_{\rm photo}|<0.2$.

In general, it is clear that the $f=1$ model tends to overestimate the EL fluxes. 
In addition, we confirm apparent trends by further comparing two panels of interest: first at fixed line fluxes, both for [OII] (left panels) and H$\alpha$ (right panels), the $f=1$ model more overpredicts the EL fluxes at lower redshift. 
For instance, at $\log_{10}(F^{\rm obs}_{\rm OII}/{\rm erg\,s^{-1}\, cm^{-2}})=-15.8$, $F^{\rm model}_{\rm OII}/F^{\rm obs}_{\rm OII}=1.73\pm 0.82$ for zCOSMOS at $0.48<z<1.54$. 
Secondly, at a similar redshift, the $f=1$ model tends to overestimate the EL fluxes at a shorter wavelength. 
At $z\sim 1$, zCOSMOS [OII] gives as large as $F^{\rm model}_{\rm OII}/F^{\rm obs}_{\rm OII}=1.73$ in the flux-complete range, while 3D-HST H$\alpha$ gives $F^{\rm model}_{\rm H\alpha}/F^{\rm obs}_{\rm H\alpha}=1.26$.\par

In Figure~\ref{fig:EBV}, we also show the results with $f=0.44$ (square points with dashed curves) which is consistent with the measurement of the dust attenuation at $z=0$ in \citet[][]{Calzetti:1994aa}.
It is apparent that the model with $f=0.44$ attenuates the EL fluxes too aggressively and generally underpredicts the EL fluxes except for zCOSMOS H$\alpha$ at $0<z<0.46$. 
For zCOSMOS H$\alpha$, the small difference between $f=0.44$ and $f=1$ cases implies that modeling the dust attenuation in star-forming nebulae is not so important to predict H$\alpha$ for the average population at $z\sim 0$. 

To reconcile these trends in the simplest $f=1$ and $f=0.44$ models, we argue that \textit{the redshift-dependent factor in the stellar-to-nebular extinction ratio}, i.e., $f(z)$ in Eq.~(\ref{eq:EBVneb}) can be a reasonable candidate. 
To confirm this further, we divide the predictions of the $f=1$ model into ones binned with the $E_{\rm star}(B-V)$ values in Figure~\ref{fig:EBV}. 
Solid lines with different colors correspond to different values of $E_{\rm star}(B-V)$. 
Note that, since the color excess is not so well constrained, we choose the grid of the color excess parameter with $\Delta E_{\rm star}(B-V)=0.1$, following \citet{Laigle:2016fr}. 
Although statistically insignificant, the predictions for galaxies with larger $E_{\rm star}(B-V)$ tend to systematically more overestimate the EL fluxes.\par 

A complication here is that the amount of the dust attenuation necessary to match the observed EL fluxes cannot be directly read from Figure~\ref{fig:EBV}.  
One should keep in mind that we simultaneously fit the model with ELs to the multi-band photometry. 
Therefore, the resultant values of the SED fitting such as the stellar mass and the color excess are also affected by the choice of different models of $f(z)$. 
Here we assume a linear evolution of $f(z)$ just for simplicity, and find that the following functional form works well, as we present in the next subsection: 
%%%%%
\begin{equation}
f(z) = 
\left\{
\begin{array}{ll}
0.44 + 0.2z & (z \le 2.8)\\
1 & (z > 2.8)
\end{array}
\right., 
\label{eq:f(z)}
\end{equation}
%%%%%
where we enforce $f$ to be one at $z>2.8$, since $f>1$ is not physically plausible.
In Figure~\ref{fig:f_z}, we compare Eq.~(\ref{eq:f(z)}) (solid line) with other results in the literature. 
As mentioned above, observations show that $f=0.44$ at $z=0$ and tend to become larger at higher redshift. 
Our finding, Eq.~(\ref{eq:f(z)}), is consistent with this trend, although there are discrepancies in detail.
Since it is under debate the detailed dependence of $f$ measurement on galaxy sample and methodology, we do not discuss further the differences. 
We also compare Eq.~(\ref{eq:f(z)}) with the result in \citet[][]{Izquierdo-Villalba:2019aa} (dashed line). 
In \citet[][]{Izquierdo-Villalba:2019aa}, they calibrated the redshift evolution of $f$ such that their prediction of the EL fluxes in their semi-analytic simulation becomes consistent with observed luminosity functions. 
Note that, although $f$ in \citet[][]{Izquierdo-Villalba:2019aa} differs for each single galaxy depending on its metallicity and inclination angle etc., we here normalize it so that $f=0.44$ at $z=0$, and focus on the difference in the evolution. Even though the two curves have been derived independently, they are both consistent, with the same trend of higher $f$ values at higher redshifts.
%Both results becomes larger consistently at larger redshift, although our linear evolution evolves less strongly and does not exceed $f=1$ in a redshift range of interest, $z\lesssim 2.5$.

%==============================%
\subsection{Results}
\label{subsec:results}
%==============================%

%~~~~~~~~~~~~~~~~~~~~~~~~~~~~~~%
\begin{figure*}
\begin{center}
\includegraphics[width=1.\textwidth]{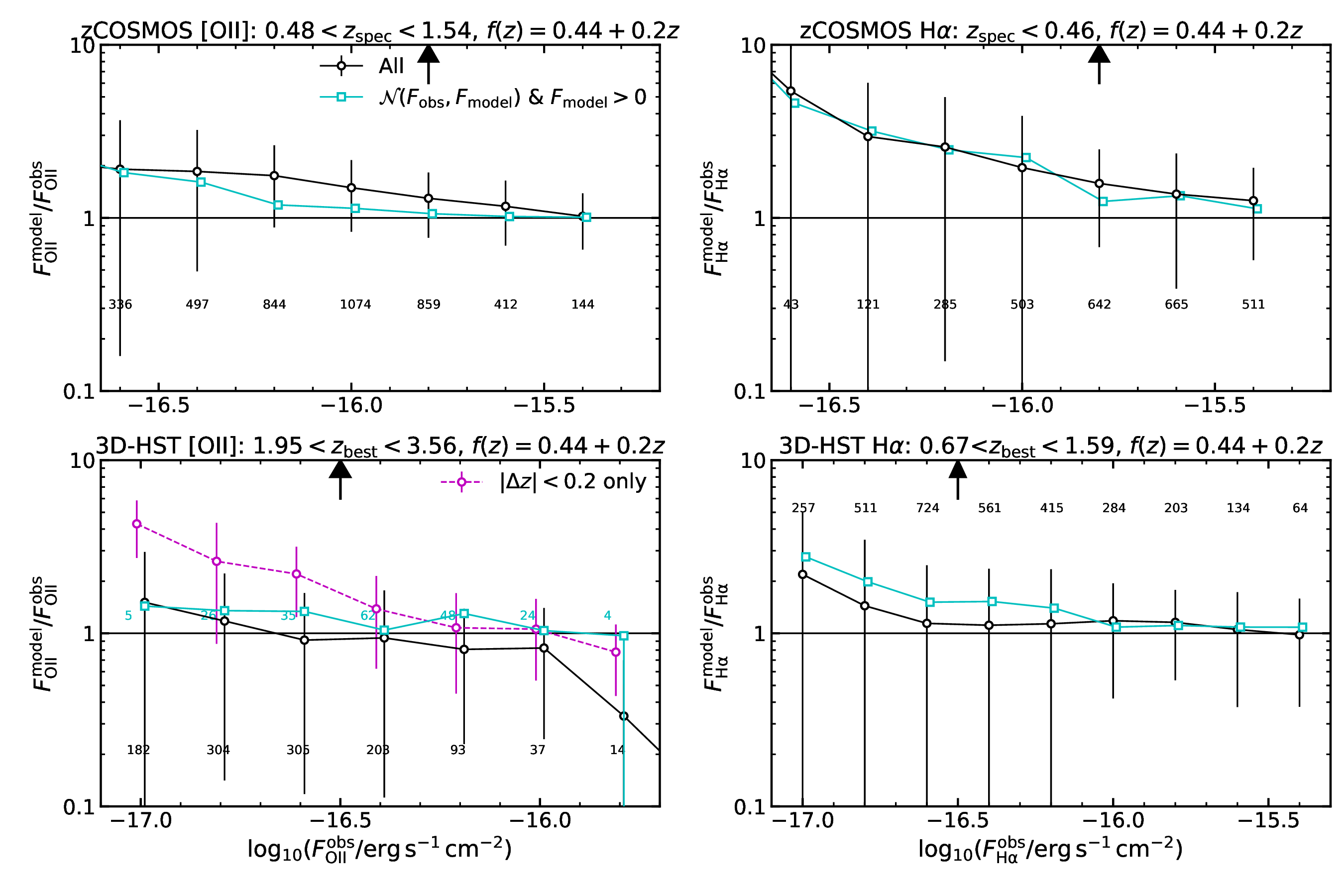}
\caption{
  \label{fig:flux_fz}
  The performance of our fiducial model with $f(z)=0.44+0.2z$.
  The structure of the figure is similar to Figure~\ref{fig:EBV}, except that we now use all galaxies at $1.95<z_{\rm best}<3.56$ for the 3D-HST [OII] case (\textit{bottom left} panel). 
  For reference, we also show the result for the restricted sample (\texttt{z\_best\_s==2} and $|\Delta z|=|z_{\rm best}-z_{\rm photo}|<0.2$) as in Figure~\ref{fig:EBV} (\textit{magenta points with a dashed line}).
  In addition, we obtain the cyan curve by measuring the mean from random variables as $F_{\rm model}$ which follow a normal distribution, $\mathcal{N}(\mu,\sigma)$, with $\mu=F_{\rm obs}$ and $\sigma=\sigma(F_{\rm model})$ but by limiting to $F_{\rm model}>0$.
}
\end{center}
\end{figure*}
%~~~~~~~~~~~~~~~~~~~~~~~~~~~~~~%

%==============================%
\subsubsection{H$\alpha$ and [OII]}
\label{subsubsec:flux_ha_o2}
%==============================%
Figure~\ref{fig:flux_fz} shows one of the main results of this paper. 
We can see in that figure that our fiducial model with $f(z)=0.44+0.2z$ performs well for both H$\alpha$ and [OII] within a factor of two 
beyond our completeness limits (arrows) in all redshift ranges.
Despite of this success, we also confirm that our model tends to more overestimate the EL fluxes in smaller flux ranges. 
%\textcolor{red}{Note that we supplement separately the impact of the EL measurements on the comparison in Appendix~\ref{appendix:impact_of_SN}.}
Even though such a failure could attribute to various assumptions and simplifications in our modeling approach (see Appendix~\ref{appendix:impact_of_SN}), let us discuss some issues which we have identified.  

As mentioned above, the standard deviations shown in Figure~\ref{fig:flux_fz} include naturally contributions from any kind of uncertainties. 
One uncertainty, which accounts at least partly for larger errors at smaller fluxes, is the one in the SED fitting to the photometry. 
Since we do not allow negative modeled flux, the distribution of $F_{\rm model}$ should be cut off at $F_{\rm model}=0$ when the width of the distribution becomes sufficiently large. 
As a result, the mean of such a skewed distribution \textit{can be larger than the original mean}. 
To see its impact, we generate random toy-model fluxes which follow a normal distribution with its mean of $F_{\rm obs}$ and its variance of $\sigma(F_{\rm model})$, and then measure the mean and the standard deviation of $F_{\rm model}/F_{\rm obs}$, cutting at $F_{\rm model}>0$. 
The result is shown in Figure~\ref{fig:flux_fz} (cyan square data points with a solid line). 
The toy-model results are roughly consistent with the overestimation trend of our predictions. 
Thus, even if our model works well on average, large uncertainty results in a skewed distribution with the cut at $F_{\rm model}>0$ and an apparent biased result.

\begin{figure}
\begin{center}
\hspace*{-0.5cm}
\includegraphics[width=1.1\columnwidth]{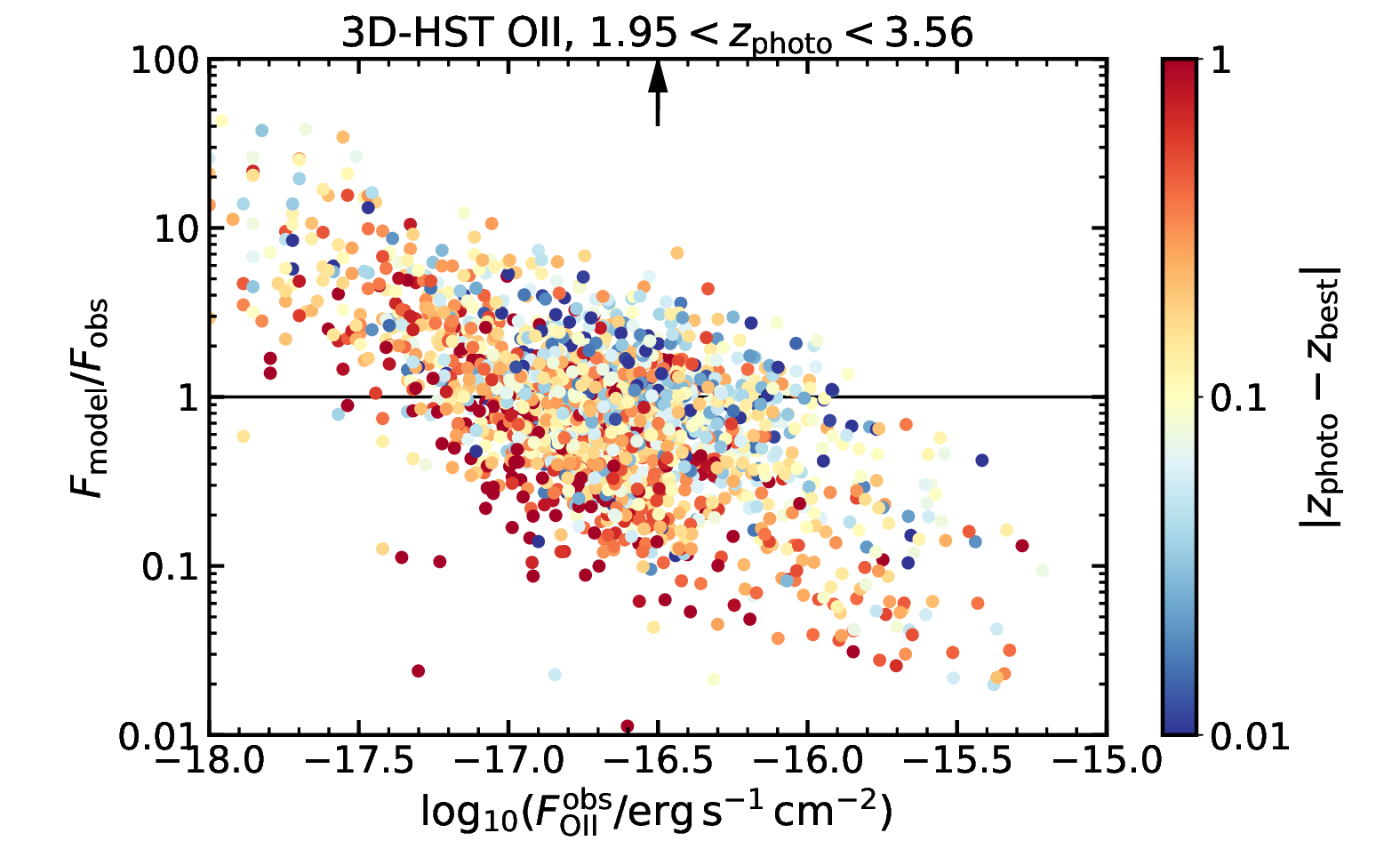}
\caption{
  \label{fig:zphoto}
  The impact of photo-z accuracy on our flux prediction 
  at $z\gtrsim 2$. 
  Each data point represents $F_{\rm model}/F_{\rm obs}$ for each 3D-HST galaxy, color-coded by the absolute difference between the photometric redshift in COSMOS2015, $z_{\rm photo}$, and the best estimate in 3D-HST catalog, $z_{\rm best}$.  
}
\end{center}
\end{figure}

In addition, we have identified that the uncertainty of photometric redshift determination impacts on our flux prediction at high redshifts of $z\gtrsim 2$. 
In the bottom-left panel of Figure~\ref{fig:flux_fz}, we have already presented that the result looks more consistent with unity, if we use the restricted sample of galaxies whose photometric redshifts are relatively close to spectroscopic ones.
We do not confirm such a drastic impact for the galaxies at lower redshift, $z\lesssim 2$.
We visualize the impact of photometric redshift more directly in Figure~\ref{fig:zphoto}. 
From Figure~\ref{fig:zphoto}, it is even clearer that our prediction tends to underestimate when we have a relatively large discrepancy between two redshift estimates, $|z_{\rm best}-z_{\rm photo}|\gtrsim 0.1$. 
Notice that, as we mentioned before, $z_{\rm best}$ in 3D-HST does not always correspond to a spectroscopically measured one. 
Since we do not know true redshifts of the entire COSMOS photometric galaxy sample, this is one of the fundamental limitations of our flux modeling at $z\gtrsim 2$. 

Finally, let us briefly comment on the consistency of derived galaxy properties between this work and \citet[][]{Laigle:2016fr} to see the impact of the EL model on the derived galaxy properties. 
We compare the stellar mass and specific SFR (sSFR). We find that our stellar mass estimate is robust against the EL modeling: the two estimates are consistent within 0.05 dex in the mean with the scatter of 0.1-0.2 dex, depending on the stellar mass.
The sSFR is more sensitive to the EL modeling. The values in \citet[][]{Laigle:2016fr} tend to be larger by 0.6-0.8 dex at $\log_{10}({\rm sSFR/yr^{-1}})<-11$, while two estimates at higher sSFR becomes consistent within 0.1-0.2 dex in the mean with a scatter of 0.3-0.5 dex. This is one of the reasons why our EL model do not rely on the SFR estimate as in other previous works. Note that this trend is consistent with the Horizon-AGN simulation result in \citet[][]{Laigle:2019aa}.

%~~~~~~~~~~~~~~~~~~~~~~~~~~~~~~%
%~~~~~~~~~~~~~~~~~~~~~~~~~~~~~~%

%~~~~~~~~~~~~~~~~~~~~~~~~~~~~~~%
\begin{figure*}
\begin{center}
\includegraphics[width=1.\textwidth]{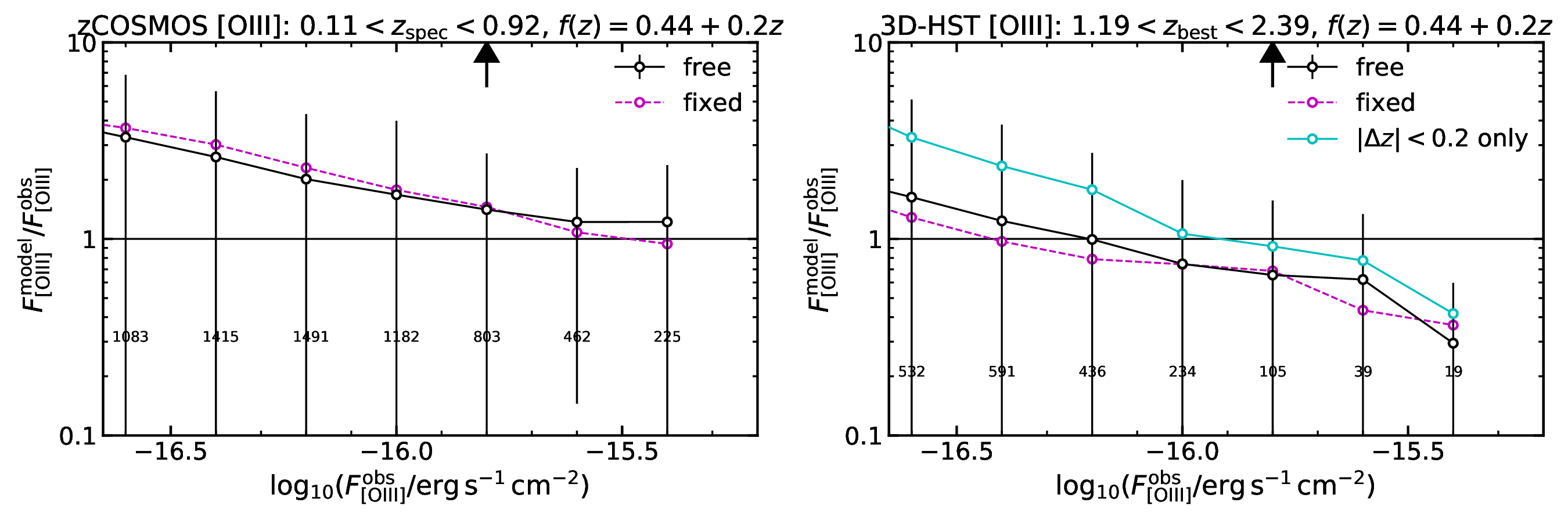}
\caption{
  \label{fig:flux_o3}
  The comparison of our predicted [OIII] fluxes with the spectroscopic measurements in zCOSMOS (\textit{left}) and 3D-HST (\textit{right}). 
  \textit{Magenta dashed} line is the result when we fix the line ratio as ${\rm [OIII]/H}\beta=4.1$, while \textit{black solid} line is the case when we let the line ratio \textit{free}. 
  As in Figure~\ref{fig:flux_fz}, we also show the result for the restricted sample (\texttt{z\_best\_s==2} and $|\Delta z|=|z_{\rm best}-z_{\rm photo}|<0.2$) (\textit{cyan solid} line).
  }
\end{center}
\end{figure*}
%~~~~~~~~~~~~~~~~~~~~~~~~~~~~~~%

%~~~~~~~~~~~~~~~~~~~~~~~~~~~~~~%
\begin{figure}
\begin{center}
\hspace*{-0.7cm}
\includegraphics[width=\columnwidth]{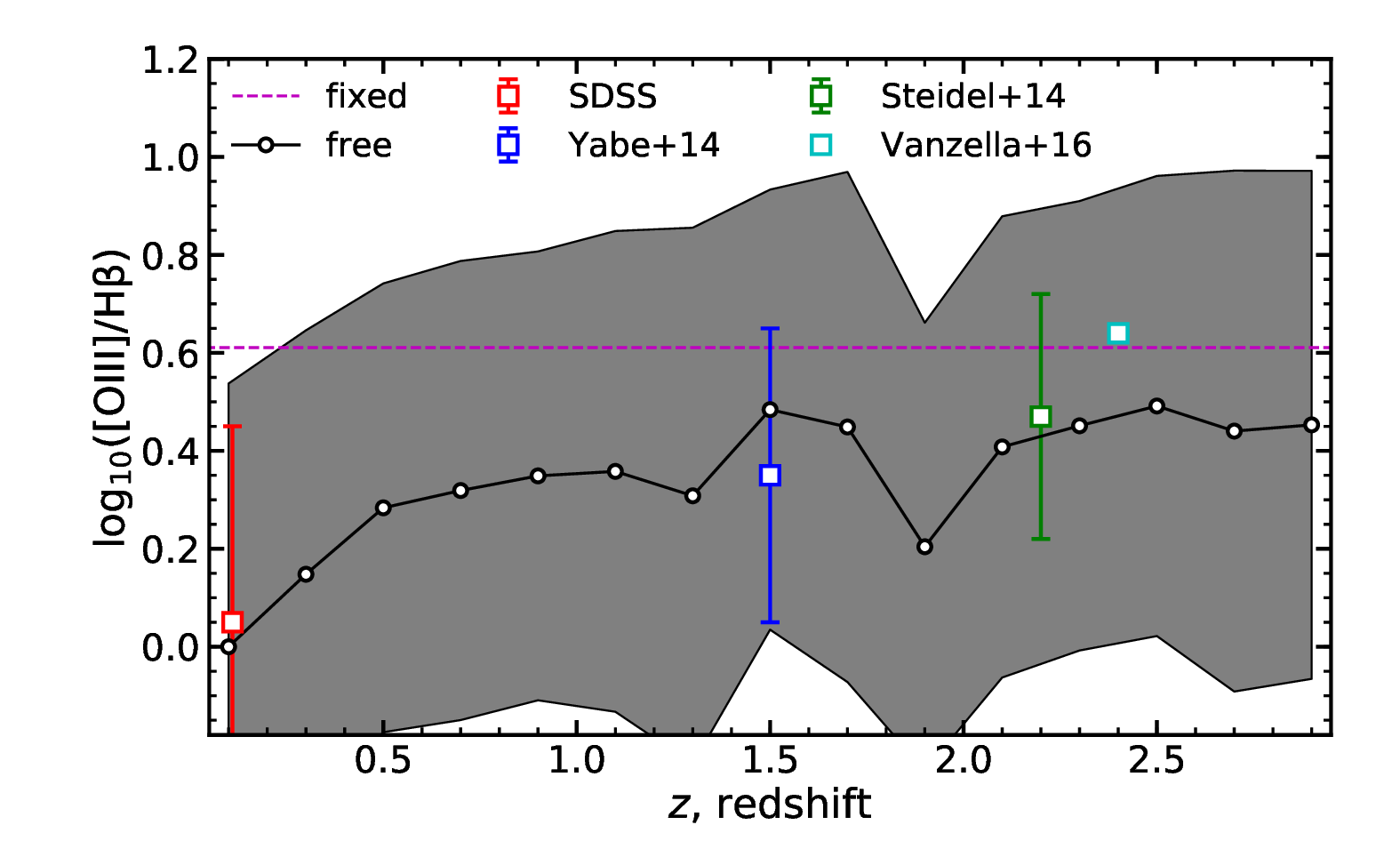}
\caption{
  \label{fig:o3hbeta}
  The redshift evolution of ${\rm [OIII]/H}\beta$. Our default model fixes it with 4.1 ({\it dashed magenta} line), while we show the result with the case of {\it free} line ratio in our SED fitting ({\it black solid} line). 
  The gray region refers to 1-$\sigma$ uncertainty. 
  For reference, we compare the direct measurements from observations in the literature; \citet[][]{Yabe:2012aa,Shapley:2005bb,Steidel:2014xx,Vanzella:2016aa}.
}
\end{center}
\end{figure}
%~~~~~~~~~~~~~~~~~~~~~~~~~~~~~~%

%==============================%
\subsubsection{[OIII]}
\label{subsubsec:flux_o3}
%==============================%
Even though our primary focus is the prediction of the [OII] and H$\alpha$ fluxes, let us briefly present the result for [OIII].
We do not use the [OIII] for the dust calibration, because of possible contribution of AGNs to the line luminosity, especially at lower sSFR \citep[e.g.,][]{Trump:2015aa}. Moreover, [OIII] is not a good indicator of SFR because it is more sensitive to ISM condition, while [OII] is more commonly used because the excitation is sufficiently well-behaved to be calibrated \citep[e.g.][]{Kennicutt:1998ab}.
In Figure~\ref{fig:flux_o3}, we show the comparison of our prediction with the spectroscopic measurements in zCOSMOS and 3D-HST. 
We note that we fix the line ratio with ${\rm [OIII]/H}\beta=4.1$ for this result. 
The result looks very similar to the [OII] case. 
Our model tends to overestimate the [OIII] fluxes at smaller fluxes, and the impact of $z_{\rm photo}$ accuracy is not negligible at high redshift range in 3D-HST. 
In general, our model works well within a factor of two beyond the completeness limit. 
However, it has been recently pointed out observationally that the line ratio, ${\rm [OIII]}/{\rm H\beta}$., evolves with redshift on average \citep[e.g.,][]{Yabe:2012aa, Kewley:2013aa}. Motivated by the fact that there could be residual information on the [OIII] EL in the intermediate wavelength ranges of the photometry, we redo our fitting by letting the line ratio, ${\rm [OIII]}/{\rm H\beta}$, be free along a coarse grid, as we described above. 
This result is also shown with black solid lines in Figure~\ref{fig:flux_o3} where we do not confirm a drastic improvement in this comparison. 
Interestingly, however, we find that this result tends to recover the trend of the increasing line ratio at high redshift, as we show in Figure~\ref{fig:o3hbeta}. 
It is not surprising to obtain such a large uncertainty, since our fitting relies only on photometric data.

%==============================%
%==============================%
\section{H\texorpdfstring{$\alpha$}{} and [OII] luminosity functions}
\label{sec:lf}
%==============================%
%==============================%

The catalog of estimated emission-line fluxes for COSMOS2015 galaxies uniquely allows us to study emission-line luminosity functions over a contiguous area of $1.38$ deg$^2$ and up to redshifts of $z=2.5$. 
In this section we present the measurements and modelling of the H$\alpha$ and [OII] luminosity functions at different redshifts in the range $0.3<z<2.5$.

%==============================%
\subsection{Estimation of the luminosity function}
\label{subsec:lf_estimate}
%==============================%

Accurate estimation of the luminosity function at different redshifts necessitates accounting for individual galaxy luminosity and redshift uncertainties, particularly when those quantities are based on photometric information. The photometric redshift uncertainty affects the estimation of the luminosity at a given flux, when fluxes are converted to luminosities through the luminosity distance. 
Additionally, our emission-line flux estimates have an intrinsic error that further impacts the estimated luminosities. 
Although on average, our H$\alpha$ and [OII] modelled fluxes are close to be unbiased as shown in Figure~\ref{fig:flux_fz}, systematic uncertainties still remain and need to be accounted for in the analysis of the luminosity function.  The effect of flux uncertainties on the luminosity function is often referred to as \textit{the Eddington bias}. Overall, the two effects affect the estimated luminosity function in a similar fashion: they introduce a luminosity-dependent smearing.  The modelling of luminosity uncertainties is addressed in the next subsection and we concentrate in the following on the impact of redshift uncertainties.

In order to derive the luminosity function from photometric samples it is crucial to account for redshift uncertainty. Unbiased estimates can be derived from the knowledge of the redshift probability distribution function associated with each galaxy. This can be achieved using deconvolution- or convolution-based estimators \citep[e.g.][]{sheth07,sheth10}. In this work we use the $V_{\rm max}$ convolution estimator described in \citet{sheth10} for which the estimated luminosity function is given by
\begin{equation} 
\label{eq:lfpdf}
    \Phi(L)= \int dL' N(L') \frac{P(L|L')}{V_{\rm max}(L)},
\end{equation}
where $L'$ is the estimated luminosity, $N(L')$ is the number of galaxies with estimated luminosity $L'$, $P(L|L')$ is the probability of the true luminosity given the estimated luminosity of $L'$, and $V_{\rm max}$ is the maximal volume in which each galaxy is observable. In practice, Eq.~(\ref{eq:lfpdf}) can be estimated as a sum over galaxy luminosity probability distribution functions
\begin{equation} 
\label{eq:lfpdfsum}
    \Phi(L) = \sum_{L'} \frac{P(L|L')}{V_{\rm max}(L)}\Theta(L),
\end{equation}
where $\Theta(L)$ is equal to unity for $L$ in $[L-dL/2, L+dL/2]$ and null otherwise. In our analysis, galaxy $V_{\rm max}(L)$ are all set to the total analysed volume. We make this choice because of the difficulty of defining $V_{\rm max}$ for emission-line galaxies in a survey selected in near-infrared apparent magnitudes \citep{Laigle:2016fr}. As a consequence, we do not correct for the Malmqvist bias affecting the faint end of the luminosity function in the estimator. Instead, we use conservative faint limits below which data are not used in the analysis (see section~\ref{subsec:lf_evolution}).

The \texttt{LePhare} code provides output information about single galaxy redshift uncertainty. In particular, it returns for each galaxy the likelihood of the photometric data for different given true redshifts $z$, i.e. the probability $P(z'|z)$. In the latter definition, $z'$ is the estimated redshift and effectively identifies the photometric data of one galaxy. This can be turned into the probability of the redshift given the data (i.e. the posterior likelihood) by using Bayes theorem, i.e. $P(z|z') = P(z'|z) P(z)$ where $P(z)$ is the prior distribution \citep[e.g.,][]{benitez00}. 
To estimate the prior distribution $P(z)$, previous measurements of the luminosity or stellar mass function evolution in the COSMOS field \citep{Ilbert:2013kx,davidzon17} can be used, from which one can derive the expected number of galaxies as a function of redshift. 
Although this may resemble a circular problem, since the redshift evolution of the luminosity function allows the prediction of the global $P(z)$, which is needed to estimate the luminosity function in the first place, the $P(z)$ can be determined iteratively and self-consistently. 
In practice, an initial guess can be used to estimate the luminosity function at different redshifts. 
The best-fitting evolutionary model to those measurements is then used to update the prior $P(z)$. 
This cycle is repeated several times to refine the estimate of the prior $P(z)$.
We find that this converges quickly after a few iterations. 
After performing various tests, we find that the exact shape of the prior $P(z)$ is not crucial and its rigorous knowledge does not affect significantly the luminosity function estimation, at least in the case of the COSMOS data.

From the $P(z|z')$ and the observed flux of each galaxy, one can derive $P(L|L')$. 
Here we assume that the flux is the true flux and has no associated error. 
The impact of flux error is treated separately (see section~\ref{subsec:error}). 
Finally, we note that, to estimate the luminosity function in redshift intervals, one needs to set $P(L|L')$ to zero in Eq.~(\ref{eq:lfpdfsum}) for $L$ associated to $z$ outside of the redshift interval considered. 
In this way, one only takes the luminosity contribution of each galaxy in the considered redshift range.

%==============================%
\subsection{Modelling luminosity error}
\label{subsec:error}
%==============================%

One can quantify the flux modelling errors by calculating the difference between estimated fluxes and direct flux measurements from reference spectroscopic samples. The full error distribution as a function of the true luminosity needs to be assessed first. 
Once this is known, one can derive the conditional probability of the luminosity difference given the true luminosity, which can then be used as convolution kernel to reproduce the observed luminosity function. The latter is thus given by
\begin{equation} \label{eq:phiobs}
    \Phi^{\rm obs}(L) = \Phi^{\rm true}(L) * P(\Delta L|L),
\end{equation}
where $\Phi^{\rm obs}(L)$, $\Phi^{\rm true}(L)$, and $P(\Delta L|L)$ are the observed luminosity function, true luminosity function, and conditional probability distribution function of the luminosity difference $\Delta L = L^{\rm est}-L$ given the true luminosity $L$, respectively. $L^{\rm est}$ is the estimated luminosity and the symbol `*' indicates the convolution product. Effectively, Eq.~(\ref{eq:phiobs}) reduces to a convolution with a variable kernel in $L$. 
We use this forward modelling approach to account for flux modelling errors later in the luminosity function analysis.

The reference spectroscopic H$\alpha$ and [OII] measurements that we are using do not fully cover the redshift and magnitude range of the COSMOS2015 catalog. 
Moreover, some spectroscopic samples have relatively low numbers of objects and are incomplete, making the probability distribution function difficult to estimate due to the high level of noise in those cases. 
To tackle those problems, we employ the extreme deconvolution method of \citet{bovy11} that generalizes the Gaussian mixture model density estimation in the case of incomplete, heterogeneous, and noisy data. 
The method provides a model for the underlying distribution in the form of a Gaussian mixture
\begin{equation} \label{eq:gmm}
M(\boldsymbol{\theta})=\sum_{i=1}^{K} w_i G(\boldsymbol{\mu_i}, \boldsymbol{\Sigma_i}),
\end{equation} 
where $\boldsymbol{\theta}$ is the parameter vector (in our case $\boldsymbol{\theta}=(\Delta L,L)$), $w_i$ are weights, $G(\boldsymbol{\mu_i}, \boldsymbol{\Sigma_i})$ is the multivariate Gaussian with means $\boldsymbol{\mu_i}$ and covariance matrix $\boldsymbol{\Sigma_i}$, and $K$ is the number of multivariate Gaussian components. The underlying assumption is that $P(\Delta L,L)$ is a smooth function, which is reasonable. We thus apply the extreme deconvolution algorithm to estimate the joint probability distribution $P(\Delta L,L)$. 
In practice, we work with the logarithm base 10 of the luminosity and the first variable of the joint probability reduces to $\Delta \log_{10} L = \log_{10}(F^{\rm est}/F)$, where $F^{\rm est}$ and $F$ are the estimated and true fluxes, respectively.

\begin{figure}
\begin{center}
\hspace*{-0.7cm}
\includegraphics[width=0.75\columnwidth, bb= 0 0 359 628]{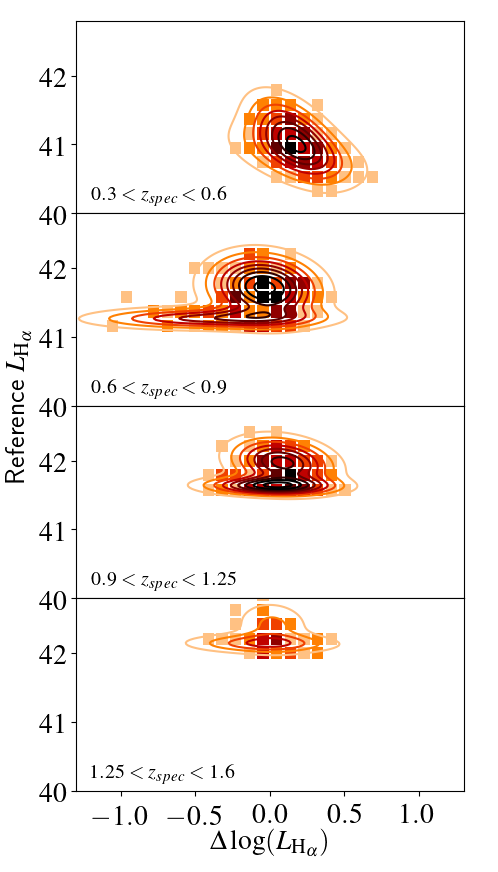}
\caption{
  \label{fig:lumhaerror}
  Joint probability distribution of $H_{\alpha}$ luminosity error and true luminosity in the spectroscopic samples zCOSMOS-Bright and 3D-HST. 
  The various panels correspond to different spectroscopic redshift intervals. 
  The colour scheme from yellow to black encodes the level of probability with arbitrary normalisation. The squares show the two-dimensional histogram of the data and the contours correspond to the best-fitting Gaussian mixture model. The colours scale similarly for both cases.
}
\end{center}
\end{figure}

In the case of H${\alpha}$, we use zCOSMOS-Bright and 3D-HST reference spectroscopic samples to cover all redshift up to $z=1.6$: H${\alpha}$ is visible in zCOSMOS-Bright at $z<0.6$ and in 3DHST sample between $z=0.6$ to $z=1.6$. We split the entire redshift range in four intervals: $0.3<z<0.6$, $0.6<z<0.9$, $0.9<z<1.25$, $1.25<z<1.6$, in which we later estimate the luminosity functions. 
The raw estimate of $P(\Delta \log_{10} L_{\rm H{\alpha}}, L_{\rm H{\alpha}})$ obtained by performing the two-dimensional histogram of the data as well as the best-fitting Gaussian mixture model obtained with the extreme deconvolution method are shown in Figure~\ref{fig:lumhaerror}. 
The best-fitting model is obtained by using a mixture of three bivariate Gaussians (i.e., with $K=3$ in Eq. \ref{eq:gmm}). 
We use the latter model as our baseline model for the conditional luminosity error in the analysis of the luminosity function.

%~~~~~~~~~~~~~~~~~~~~~~~~~~~~~~%
%~~~~~~~~~~~~~~~~~~~~~~~~~~~~~~%

\begin{figure}
\begin{center}
\hspace*{-0.7cm}
\includegraphics[width=0.75\columnwidth, bb=0 0 360 492]{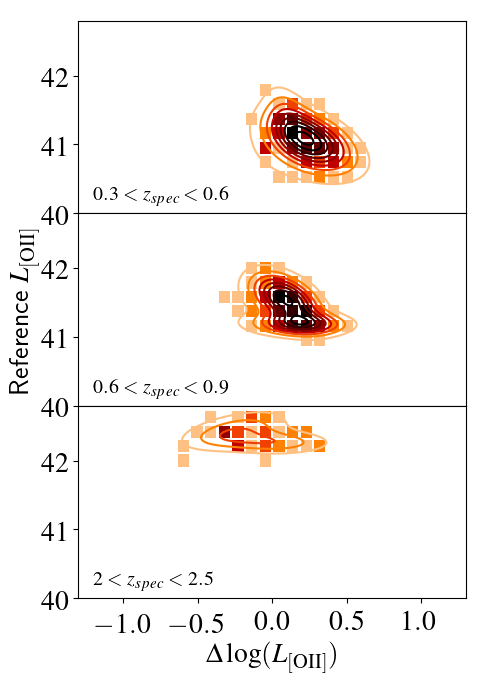}
\caption{
  \label{fig:lumo2error}
  Same as Fig. \ref{fig:lumhaerror} but for the case of [OII].
  }
\end{center}
\end{figure}

In the case of [OII], we cannot estimate the flux modelling errors over the entire COSMOS redshift range, since [OII] fluxes are not always visible in the reference spectroscopic samples at our disposal. We can use zCOSMOS at $z<1$ and 3D-HST at the highest redshifts, between $z=2$ and $z=2.5$. Similarly to for H${\alpha}$, we estimate $P(\Delta \log_{10} L_{\rm [OII]}, L_{\rm [OII]})$ using the extreme deconvolution method on the three redshift intervals: $0.3<z<0.6$, $0.6<z<0.9$, and $2<z<2.5$. The best-fitting models are obtained by using a mixture of three bivariate Gaussians and are shown in Figure~\ref{fig:lumo2error}. 
We note that, in the estimation of luminosity error distribution, we only consider spectroscopic flux measurements that have a signal-to-noise ratio above three. 

%~~~~~~~~~~~~~~~~~~~~~~~~~~~~~~%
%~~~~~~~~~~~~~~~~~~~~~~~~~~~~~~%

%==============================%
\subsection{Measurements of the H$\alpha$ and [OII] galaxy luminosity functions}
\label{subsec:lf_evolution}
%==============================%

\begin{figure*}
\begin{center}
\includegraphics[width=2\columnwidth, bb=0 0 906 555]{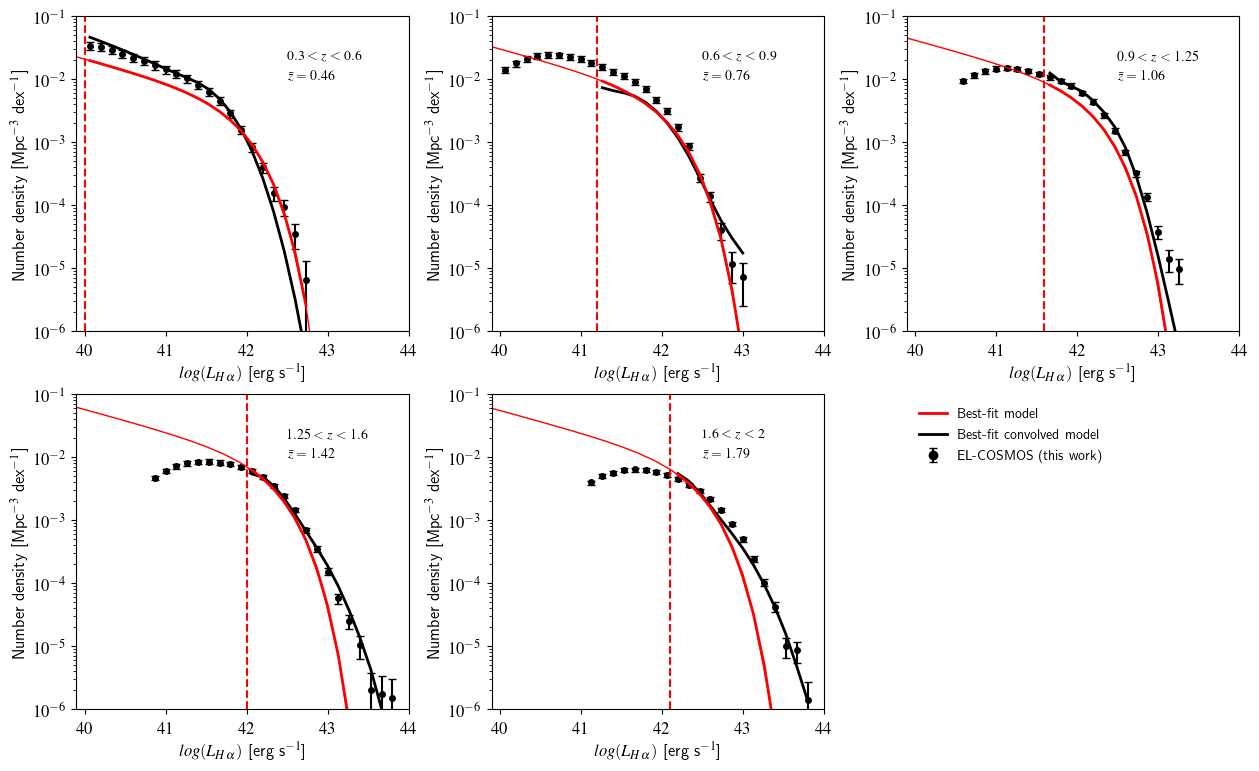}
\caption{
  \label{fig:haLF} 
  The estimated H$\alpha$ luminosity functions from the EL-COSMOS catalog in the six considered redshift intervals between $z=0.3$ and $z=2$. The effective redshift as defined by Eq. \ref{eq:effz} is provided in each panel. The black solid curves show our best-fitting models including the convolution by the luminosity error distribution, while red solid curves correspond to the associated intrinsic luminosity function models. The vertical dashed lines show the luminosity completeness limits considered in the fit.
}
\end{center}
\end{figure*}

We measure the $H_{\alpha}$ luminosity function in five redshift intervals: $0.3<z<0.6$, $0.6<z<0.9$, $0.9<z<1.25$, $1.25<z<1.6$, and $1.6<z<2$. Those measurements are shown in Figure~\ref{fig:haLF}. 
Error bars include both Poisson and sample variance errors and the two contributions are summed up in quadrature. 
The sample variance contribution in each redshift interval is estimated using the method of \citet{moster11} and assuming a Planck 2015 matter power spectrum and the galaxy linear biases given in \citet[][]{Orsi:2014aa}. 
For simplicity, we ignore off-diagonal components in the covariance matrix, whose impact is not important at the bright end \citep{Smith:2012cc}. 
We define conservative completeness limit for each redshift interval at faint luminosities. These are $\log_{10}(L_{\rm lim}/{\rm erg\, s}^{-1})=(40,41.1,41.5,42,42.3)$ for the different redshift intervals, respectively. 
Note that the completeness limits adopted here are different from the ones in Figure~\ref{fig:completeness} but are consistent.
It is clear from Figure~\ref{fig:haLF} that our data are mostly sensitive to the bright end of the luminosity function, except in the low-redshift interval at $0.3<z<0.6$. 
In the case of [OII], we restrict the analysis to the three redshift intervals:  $0.3<z<0.6$,  $0.6<z<0.9$, and $2<z<2.5$, where the luminosity error modelling can be performed. 
We adopt the completeness limits of $\log_{10}(L_{\rm lim}/{\rm erg\,s}^{-1})=(40, 41, 42)$ for the  $0.3<z<0.6$,  $0.6<z<0.9$, and $2<z<2.5$ intervals, respectively. 
The [OII] luminosity function measurements are presented in Figure~\ref{fig:o2LF}.

%~~~~~~~~~~~~~~~~~~~~~~~~~~~~~~%
%~~~~~~~~~~~~~~~~~~~~~~~~~~~~~~%

Now, let us model the observed luminosity functions with a time-evolving empirical model, including the effect of luminosity errors originated from flux modelling. 
We assume a \citet{schechter76} function with varying $\Phi_*$ and $L_*$ with redshift. 
We model those redshift variations with specific dependencies in powers of $1+z$. Similarly as in \citet[][]{geach10}, we use a double power law model for the normalization parameter $\phi^*(z)$ to gain in flexibility. 
After several tests with different parameterizations, we find that this choice best reproduce the observed evolution with redshift of the luminosity function in our data for both H$\alpha$ and [OII]. 
The model for the observed luminosity function at redshift $z$ is
\begin{equation} \label{eq:phifullmodel}
    \Phi^{\rm obs}(L|z) = \Phi^{\rm true}(L|z) * P(\Delta L|L,z),
\end{equation}
where
\begin{equation} \label{eq:phischeshter}
    \Phi^{\rm true}(L|z) dL = \Phi_*(z) \left( \frac{L}{L_{*,0}(1+z)^\beta} \right)^\alpha e^{-\frac{L}{L_{*,0}(1+z)^\beta}} \frac{dL}{L_{*,0}(1+z)^\beta},
\end{equation}
and
\begin{equation}
    \Phi_*(z) = \left\{
    \begin{array}{ll}
        \Phi_{*,0}(1+z)^\gamma & \mbox{for } z<z_{\rm pivot}\\
        \Phi_{*,0}(1+z_{\rm pivot})^{\gamma+\epsilon}(1+z)^{-\epsilon} & \mbox{for } z>z_{\rm pivot}.
    \end{array}
    \right.
\end{equation}
The conditional probability $P(\Delta L|L,z)$ in Eq.~(\ref{eq:phifullmodel}) corresponds to the best-fitting Gaussian mixture models in the redshift intervals presented in section~\ref{subsec:error}. Overall, the model has 7 free parameters: $\Phi_{*,0}$, $L_{*,0}$, $\alpha$, $\beta$, $\gamma$, $\epsilon$, and $z_{\rm pivot}$. 
However, our measurements do not allow us to constrain all of these with the same precision. 
In particular, the faint-end slope index $\alpha$ cannot be well constrained with our data alone because of our bright completeness limits except at $0.3<z<0.6$. 
We therefore fix $\alpha=-1.35$ ($\alpha=-1.25$) in the case of H$\alpha$ ([OII]). 
We choose those values as they best reproduce our low-redshift faint-end luminosity function measurements. 
Those are very close to the usually assumed or measured values in the literature \citep{pozzetti16,Comparat:2016zz}. We do not consider any evolution with redshift for this parameter in our model.

In the case of H$\alpha$, the 3D-HST spectroscopic sample does not cover the last redshift interval considered here and we lack an estimate of the luminosity error distribution in that interval. 
We find however that the observed luminosity function is quite similar to that in the previous redshift interval at $1.25<z<1.6$, particularly in the way it drops at the highest luminosities. 
The apparent excess of bright galaxies is the reflection of the convolution with the luminosity error. Since we do not expect a strong evolution in the shape of the luminosity function in the last two bins and given the similarity in the shape of the luminosity function, we assumed the luminosity error distribution to be quite similar. 
We therefore use the same $P(\Delta L|L, z)$ at $1.25<z<1.6$ and $1.6<z<2$.

%\begin{table*}
%    \centering
%    \begin{tabular}{cccccccc}
%        \hline
%        Emission line & $\log_{10} \Phi_{*,0}$ & $\gamma$ & $\epsilon$ & $z_{\rm pivot}$ & $\log_{10} L_{*,0}$ & $\beta$ & $\alpha$ (fixed) \\
%        \hline
%        ${\rm H\alpha}$ & 2.92$\pm$0.03 & 1.30$\pm$0.12 & 1.86$\pm$1.40 & 1.53$\pm$0.12 & 
%        41.59$\pm$0.03 &  1.91$\pm$0.08 & -1.35 \\
%        ${\rm [OII]}$ & 1.89$\pm$0.04 & -1.96$\pm$0.18 & -2.48$\pm$0.21 & 1.00$\pm$0.02 & 40.73$\pm$0.02 &  2.61$\pm$0.08 & -1.25 \\
%        \hline
%        \hline
%    \end{tabular}
%    \caption{H$_{\rm \alpha}$ and [OII] luminosity function parameter central values and associated $68\%$ marginal uncertainties. $\Phi_{*,0}$ and $L_{*,0}$ are in units of Mpc$^{-3}$ and erg s$^{-1}$ respectively.}
%    \label{tab:lfparams}
%\end{table*}

\begin{table*}
    \centering
    \begin{tabular}{cccccccc}
        \hline
        Emission line & $\log_{10} \Phi_{*,0}$ & $\gamma$ & $\epsilon$ & $z_{\rm pivot}$ & $\log_{10} L_{*,0}$ & $\beta$ & $\alpha$ (fixed) \\
        \hline
        ${\rm H\alpha}$ & -2.92$\pm$0.03 & 1.30$\pm$0.12 & 1.86$\pm$1.40 & 1.53$\pm$0.12 & 
        41.59$\pm$0.03 &  1.91$\pm$0.08 & -1.35 \\
        ${\rm [OII]}$ & -1.89$\pm$0.04 & -1.96$\pm$0.18 & -2.48$\pm$0.21 & 1.00$\pm$0.02 & 40.73$\pm$0.02 &  2.61$\pm$0.08 & -1.25 \\
        \hline
        \hline
    \end{tabular}
    \caption{H$_{\rm \alpha}$ and [OII] luminosity function parameter central values and associated $68\%$ marginal uncertainties. $\Phi_{*,0}$ and $L_{*,0}$ are in units of Mpc$^{-3}$ and erg s$^{-1}$ respectively.}
    \label{tab:lfparams}
\end{table*}

\begin{figure}
\begin{center}
\includegraphics[width=\columnwidth, bb=0 0 857 857]{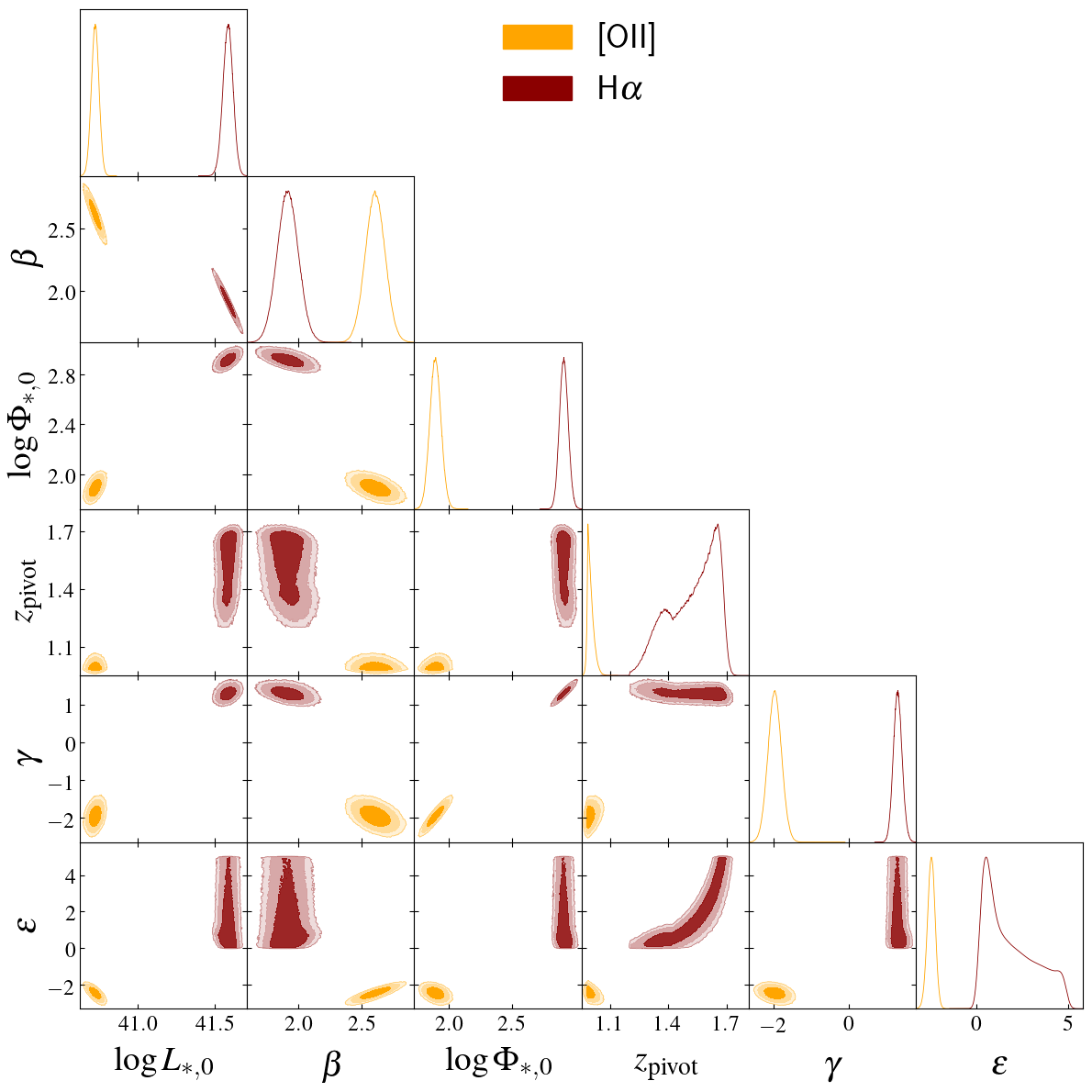}
\caption{
  \label{fig:like} 
  The posterior probability distribution function of the luminosity function parameters. The constraints for the H$\alpha$ and [OII] parameters are show in orange and red respectively. The contours correspond to $68\%$, $95\%$, and $99\%$ confidence levels for the various pairs of parameters.
}
\end{center}
\end{figure}

\begin{figure*}
\begin{center}
\includegraphics[width=2\columnwidth, bb=0 0 905 557]{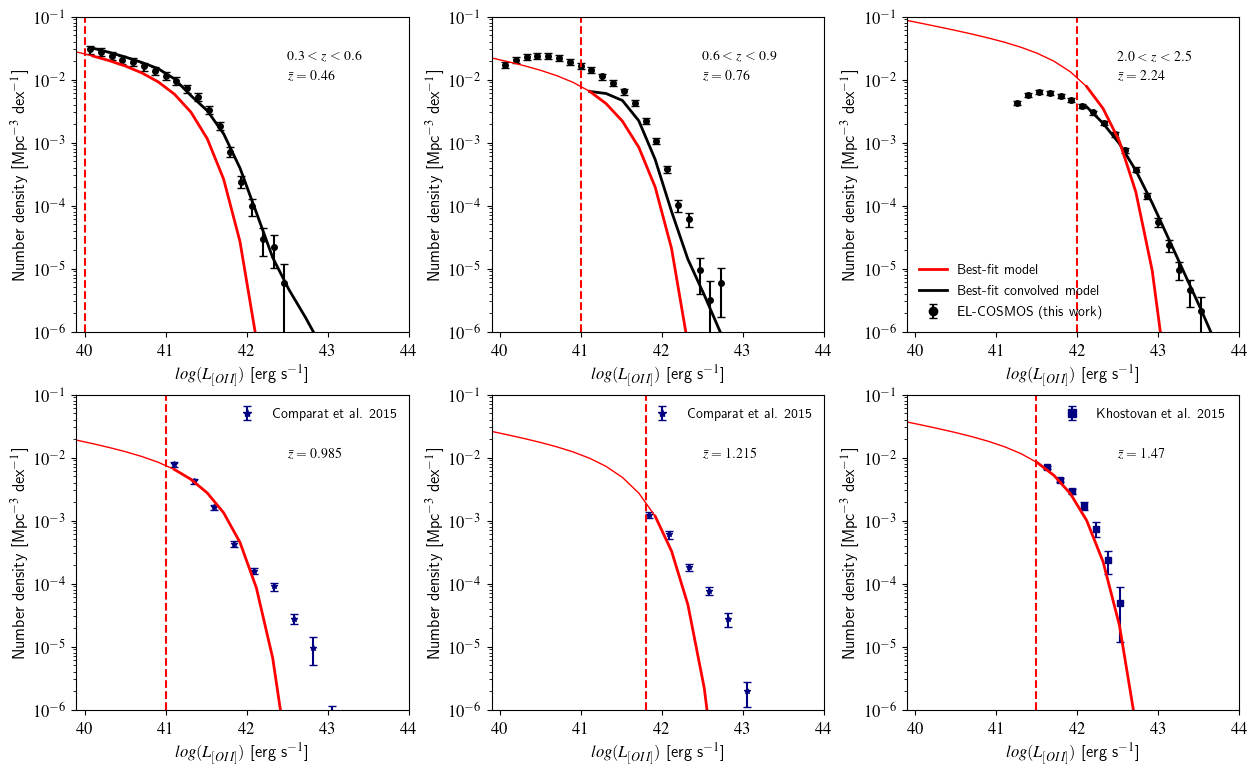}
\caption{
  \label{fig:o2LF} 
  The estimated [OII] luminosity functions from our Emission-Line COSMOS catalog in the three considered redshift intervals are presented in the top panels. The bottom panels show the external measurements from \citet[][]{Comparat:2015aa} and \citet[][]{Khostovan:2015aa}, which we also include in our analysis. The effective redshift as defined by Eq. \ref{eq:effz} is provided in each panel. The black solid curves show our best-fitting models including the convolution by the luminosity error distribution (only for EL-COSMOS measurements), while red solid curves correspond to the associated intrinsic luminosity function models. The vertical dashed lines show the luminosity completeness limits considered in the fit.
}
\end{center}
\end{figure*}

We perform a likelihood analysis of the combined observed luminosity functions in all redshift intervals, considering all luminosity bins above the luminosity completeness limits defined previously. For this purpose, we use the Monte Carlo Markov Chain ensemble sampler {\sc emcee} \citep{emcee}.  
We define the effective redshift associated to each redshift interval as the mean of the summed  individual galaxy $P(z|z')$ in the subsamples, i.e.  
\begin{equation} \label{eq:effz}
    \bar{z}=\frac{\int \left(\sum_{z'} P(z|z')\right) z dz}{\int \sum_{z'} P(z|z') dz}
\end{equation}
The best-fitting convolved model is shown with the black curve in Figures.~\ref{fig:haLF} and \ref{fig:o2LF}, while the underlying error-free model is shown in red. 
One can see in those figures that our models reproduce well the observed evolution of the  luminosity function with the exception of the interval $0.6<z<0.9$. 
In the latter interval, the observed luminosity function is significantly above the best-fitting model. 
This can be explained by invoking sample variance, the fact that in this particular volume there is potentially a high overdensity that locally boosts the number of H$\alpha$ emitters. In fact, such effect has already been evidenced in previous analyses in the COSMOS field in this particular redshift interval with the discovery of an overdense structure in form of a wall at $z=0.73$ \citep[][]{cassata07,delatorre10,iovino16}. 
The posterior likelihood contours for H$\alpha$ and [OII] luminosity function parameters are presented in Figure~\ref{fig:like}. 
The reduced $\chi_{\rm min}^2$ are $\chi_{\rm min}^2/{\rm dof}=309.1/69\simeq4.5$ and $\chi_{\rm min}^2/{\rm dof}=441.2/62\simeq7.1$ for H$\alpha$ and [OII] luminosity functions fits respectively.

%~~~~~~~~~~~~~~~~~~~~~~~~~~~~~~%
%~~~~~~~~~~~~~~~~~~~~~~~~~~~~~~%

In the case of [OII], our luminosity function measurements are restricted to three redshift intervals between $z=0.3$ and $z=2.5$ with a gap at $0.9<z<2$. 
This makes it unrealistic to constrain the entire [OII] luminosity function evolution at $0.3<z<2.5$ with our data alone. 
We therefore add external [OII] luminosity function measurements at $0.9<z<2$ in the fit. 
We include three of the most complete and accurate luminosity function measurements. 
Those are $z=0.95$ and $1.22$ measurements from \citet{Comparat:2015aa} and the $z=1.47$ measurement from \citet{Khostovan:2015aa}, which are shown in the lower panels of Figure~\ref{fig:o2LF}. 
%~~~~~~~~~~~~~~~~~~~~~~~~~~~~~~%
%~~~~~~~~~~~~~~~~~~~~~~~~~~~~~~%
We can see in the figure that we are able to fit reasonably well with our model the luminosity functions at all redshifts, including those coming from external datasets. 
We note that in the regime of bright luminosities, previous measurements tend to show an excess from a pure Schechter functional form. This can be due to contamination from AGNs in those samples, or simply the contribution from starburst galaxies, as predicted by semi-analytical models for instance \citep[e.g.,][]{Gonzalez-Perez:2018aa}. 
Given those uncertainties, we do not attempt to model this feature in the observed [OII] luminosity functions.

%==============================%
\subsection{Comparison to the literature}
\label{subsec:lf_litterature}
%==============================%

%~~~~~~~~~~~~~~~~~~~~~~~~~~~~~~%
\begin{figure*}
\begin{center}
\includegraphics[width=2\columnwidth, bb=0 0 906 556]{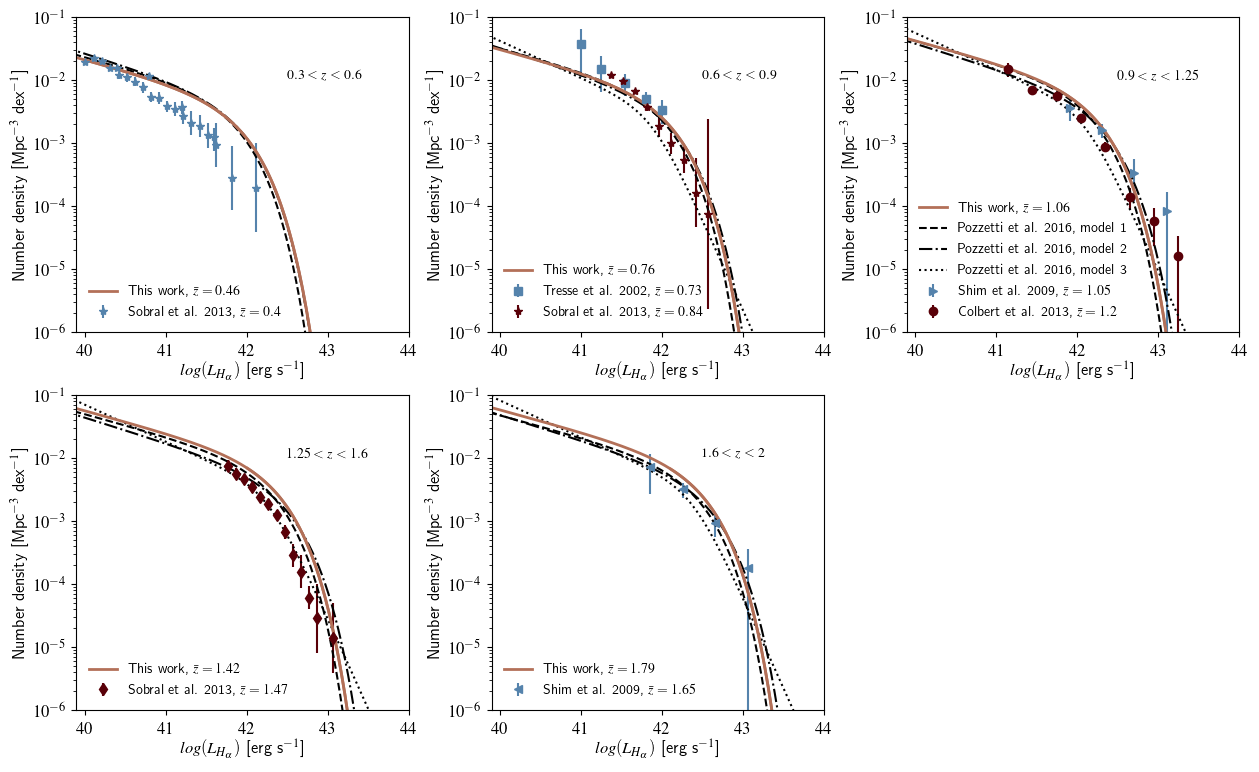}
\caption{
  \label{fig:haLFlit} 
  Comparison between our best-fit H$\alpha$ luminosity function model and previous measurements from the literature in different redshift intervals (see legends). The measurements from the literature are shown with the different points \citep{tresse02,sobral13,shim09,colbert13}, their color encoding the effective redshift. Our model is presented with the solid line and \citet{pozzetti16} models 1, 2, and 3 with the dashed, dot-dashed, and dotted lines respectively. 
}
\end{center}
\end{figure*}
%~~~~~~~~~~~~~~~~~~~~~~~~~~~~~~%

Our predicted H$\alpha$ luminosity functions are compared in Figure~\ref{fig:haLFlit} with the previous measurements compiled by \citet{pozzetti16} and whose effective redshifts fall inside our redshift intervals. 
There is an overall agreement with previous direct measurements, which nonetheless show a significant scatter at similar redshifts. 
It is important to emphasize that most of these have been obtained from small-size surveys, below one square degree, and are significantly more affected by sample variance than we are.
\citet{pozzetti16} used those measurements to model the redshift evolution of the H$\alpha$ luminosity function, defining three different evolutionary models. Their models are reported in the figure. We recall that our model has a very similar functional form as their model 1. We find a very good agreement between our best-fitting model and their model 1 and model 2 predictions. Our model tends to fall in between those two models in the bright end. The difference in number density with model 1 increases at $z>1.25$.

\citet{Valentino:2017nx} used a similar method to estimate H$\alpha$ line fluxes in the COSMOS2015 and GOODS-S fields, and from them derive H$\alpha$ luminosity function and counts. 
They restricted their analysis to the redshift range $0.9<z<1.8$ and found luminosity functions that fall in between \citet{pozzetti16}'s models 1 and 3, therefore below our predictions. 
In their work, they used the observed galaxy SED and rest-frame UV to estimate H$\alpha$ flux with \citet{Kennicutt:1998ab} relation. 
Their treatment of the dust attenuation however slightly differs from ours. 
They assumed a constant stellar-to-nebular attenuation ratio of $f=0.57$ calibrated on FMOS-COSMOS spectroscopic sample, while we used a more detailed redshift-dependent $f(z)$ calibrated on zCOSMOS and 3D-HST samples, taking advantage of the larger statistic available in those samples. 
Moreover, while they assumed a constant H$\alpha$ luminosity estimation error, we account for its redshift and intrinsic luminosity dependences. Finally, we provide a proper treatment of photometric redshift error based on the full usage of the redshift probability distribution function for each galaxy. Photometric redshift errors can have an impact on the luminosity estimation, and in turn, on the luminosity function estimation, which was neglected in \citet{Valentino:2017nx}. These methodological differences may explain the different H$\alpha$ counts (see Section \ref{sec:discussion}).

\begin{figure*}
\begin{center}
\includegraphics[width=2\columnwidth, bb=0 0 901 307]{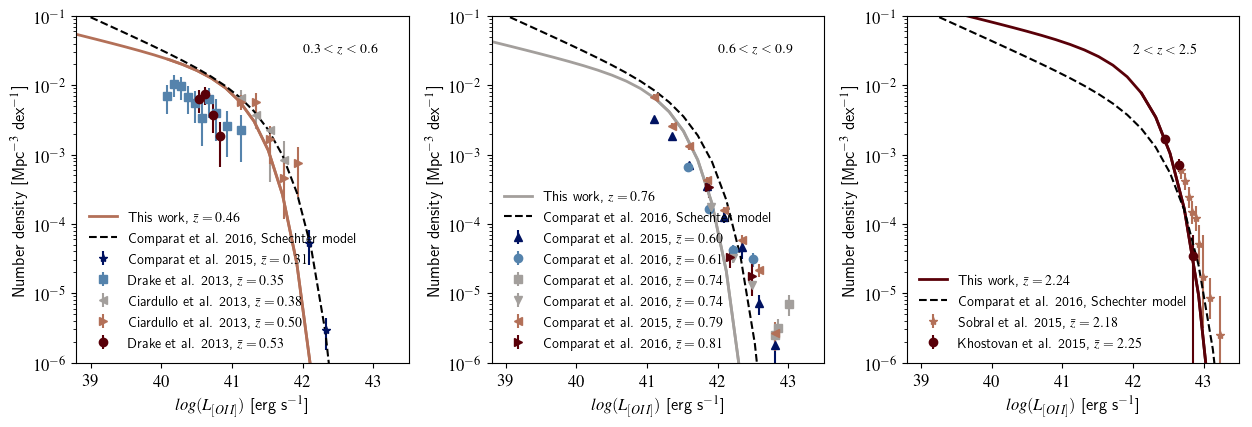}
\caption{
  \label{fig:o2LFlit}
  Comparison between our best-fit [OII] luminosity
  function model and previous measurements from the literature in different redshift intervals (see legends). The measurements from the literature are shown with the different points \citep{drake13,ciardullo13,Comparat:2015aa,Comparat:2016zz,sobral15,Khostovan:2015aa}, their color encoding the effective redshift. Our model is presented with the solid line and \citet{Comparat:2016zz} model with the dashed line for comparison.
}
\end{center}
\end{figure*}

Our predicted [OII] luminosity functions are compared to the \citet{Comparat:2016zz} compilation of previous measurements in Figure~\ref{fig:o2LFlit}. The comparison is done for measurements whose effective redshifts fall inside our redshift intervals. On can see that in each interval, there is significant scatter among measurements. Part of this comes from a significant redshift evolution of the luminosity function. Nonetheless, at similar effective redshifts we find a good agreement between our best-fit luminosity function model and previous measurements given the uncertainties. We also compare our model to the Schechter model of \citet{Comparat:2016zz}, shown with the dashed line in Figure~\ref{fig:o2LFlit}. The two models show very similar predictions above $z=2$ on the bright end, while at $z<0.9$ \citet{Comparat:2016zz} predicts more bright [OII] emitters.

%~~~~~~~~~~~~~~~~~~~~~~~~~~~~~~%
%~~~~~~~~~~~~~~~~~~~~~~~~~~~~~~%

%==============================%
\section{Prediction of H$\alpha$- and [OII]-emitter galaxy counts}
\label{sec:discussion}
%==============================%

\begin{figure}
\begin{center}
\includegraphics[width=\columnwidth, bb=0 0 505 456]{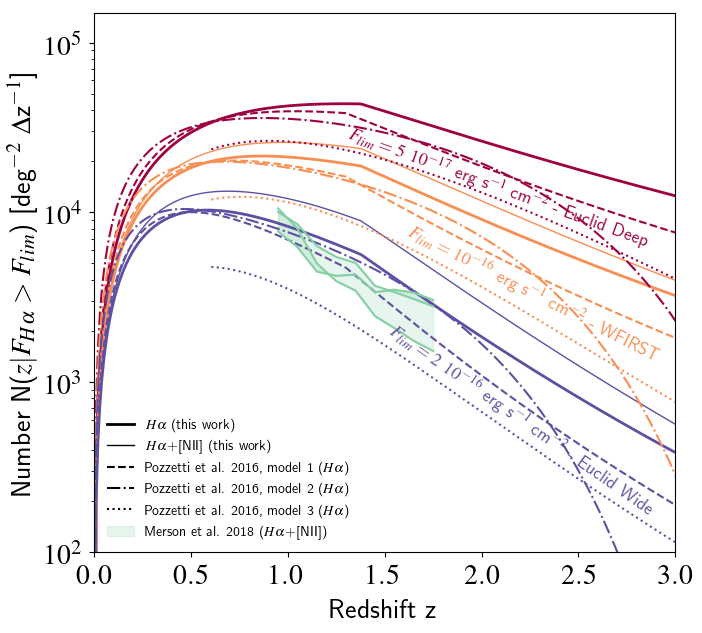}
\caption{
  \label{fig:haNz} 
  The predicted number of H$_\alpha$-emitter galaxies per deg$^2$ as a function of redshift for the three limiting fluxes (from top to bottom): $F_{\rm lim}/({\rm erg\,s^{-1}\,cm^{-2}})= 5\times10^{-17}$, $1\times 10^{-16}$, and $2\times 10^{-16}$, corresponding to the limits for Euclid Deep, WFIRST, Euclid Wide surveys respectively. Our predictions are shown with the thick solid curves, while \citet{pozzetti16} model 1 and model 3 are shown with the dashed and dotted curves respectively. We also present our predicted H$_\alpha$+[NII] redshift distributions for the two brightest flux limits with the thin solid curves. 
  The semi-analytical model predictions from \citet{Merson:2017bu} for the Euclid Wide case assuming (from top to bottom) \citet{Calzetti:2000kx}, \citet{ferrara99}, \citet{charlot00} attenuation models are also shown.
}
\end{center}
\end{figure}

We present in Figure \ref{fig:haNz} the H$\alpha$ galaxy counts as a function of redshift for different limiting fluxes: $F_{\rm lim}/({\rm erg\,s^{-1}\,cm^{-2}})= 5\times10^{-17}$, $1\times 10^{-16}$, and $2\times 10^{-16}$. 
These are compared with the predictions from \citet{pozzetti16} model 1, 2, and 3. Their models 1 and 3 are current baseline for the Euclid mission cosmological predictions \citep{Laureijs:2011aa}. We also include the expected counts for the complex H$\alpha$+[NII], since H$\alpha$ and [NII] lines will be blended at Euclid and WFIRST spectral resolutions. For this, we use the stellar mass- and redshift-dependent H$\alpha/$[NII] empirical model from \citet{faisst18}.

As expected from the luminosity function comparison in Figure~\ref{fig:haLF}, our model predicts more H$\alpha$-emitter galaxies than the most optimistic model of \citet{pozzetti16} (model 1), particularly at the highest redshifts. By integrating those redshift distributions over the redshift range where H$\alpha$ will be visible with Euclid red grism in the Euclid Wide survey, i.e. $0.9<z<1.8$, we find an expected number of galaxies at the Euclid wide survey depth of $F_{\rm H\alpha}>2 \times 10^{-16}\,{\rm erg\,s^{-1}\,cm^{-2}}$ of $6484\pm69$ galaxies per deg$^{2}$. For the H$\alpha$+[NII] complex at similar depth and redshift range, we expect $9992\pm85$ galaxies per deg$^{2}$. 
At the depth of the Euclid Deep survey, i.e. $F_{\rm H\alpha}>5 \times 10^{-17}\,{\rm erg\,s^{-1}\,cm^{-2}}$, we expect $47955\pm186$ galaxies per deg$^{2}$ in the same redshift range.
\citet{Valentino:2017nx} found for the Euclid Wide survey, and using COSMOS data, a surface density of $3796$ galaxies per deg$^{2}$ in the case of the H$\alpha$+[NII] complex. 
This is only $38\%$ of what we found and also lower than predicted by \citet{pozzetti16} model 3. 

The WFIRST survey aims at targeting H$\alpha$ emitters with grism spectroscopy at a smaller flux limit compared to Euclid Wide survey. The expected WFIRST H$\alpha$ flux limit is $10^{-16}$ erg s$^{-1}$ cm$^{-2}$ \citep{Spergel:2013aa}. The corresponding number counts predictions in the range of WFIRST grism observations, i.e. $0.5<z<1.9$, is $28600\pm144$ or $35506\pm160$ galaxies per deg$^{2}$ for the H$\alpha$+[NII] complex. Overall, our model tends to predict between $20\%$ and $34\%$ more H$\alpha$ galaxies than previously expected at those flux limits, which may improve the cosmological constraining power of Euclid and WFIRST surveys.

We also compare our H$\alpha$+[NII] counts prediction to the semi-analytical model predictions of \citet[][]{Merson:2019pp}. The latter work predicts the number counts for the Euclid Wide survey (based on \citet{pozzetti16} model 3) using three different models for dust attenuation: \citet{Calzetti:2000kx}, \citet{ferrara99}, and \citet{charlot00}. 
Those are shown in Figure~\ref{fig:haNz} and are below the expectation from our model.

%~~~~~~~~~~~~~~~~~~~~~~~~~~~~~~%
%~~~~~~~~~~~~~~~~~~~~~~~~~~~~~~%

Subaru PFS is another forthcoming galaxy redshift survey whose cosmology part aims at performing a $1400$ deg$^2$ galaxy survey at redshifts typically between $z=0.6$ and $z=2.4$ \citep{Takada:2014sf} (see also Figure~\ref{fig:zred_coverage}). 
%% I removed this, because this is already written in Introduction. 
%The spectroscopic galaxy target selection will use the deep $g$, $r$, $i$, $z$, $y$ HSC bands photometry. 
The current baseline selection is based on a $g$ magnitude and $g-r$ colour criterion, optimized to preselect galaxies above redshift one \citep{Takada:2014sf}. 
The spectral range of the spectrograph, its sensitivity, and the planned observing strategy make the survey mostly sensitive to [OII] emitters in that redshift range. 
Thus, the ability of determining galaxy redshifts will be closely related to that of detecting the [OII] emission lines in the galaxy spectra. 
The [OII] detectability can be adjusted by setting a signal-to-noise ratio (SNR) threshold that ensures a reliable galaxy redshift determination.

\begin{figure}
\begin{center}
\includegraphics[width=\columnwidth, bb=0 0 505 456]{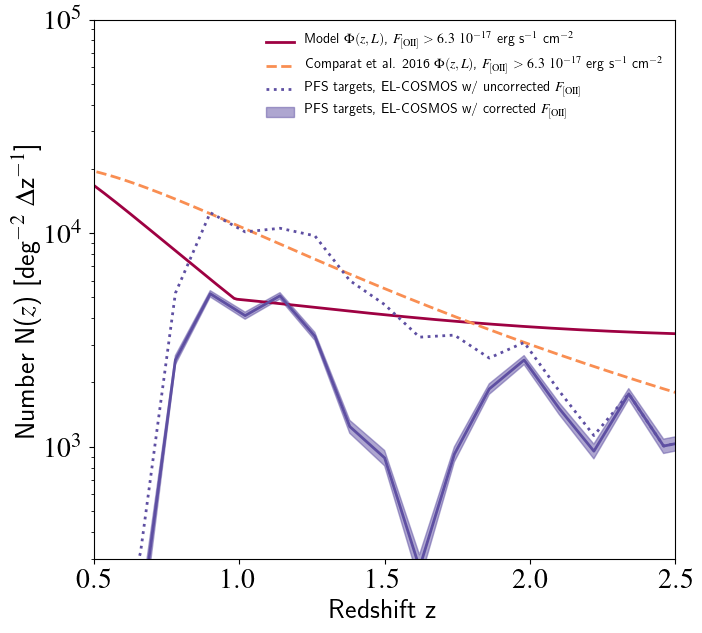}
\caption{
  \label{fig:O2Nz} 
   The predicted number of [OII]-emitter galaxies per deg$^2$ targeted in the PFS survey as a function of redshift. The {\it red solid} line is our prediction from the luminosity function when we only impose a flux limit of 6.3 $10^{-17}$ erg s$^{-1}$ cm$^{-2}$. This is compared with the prediction from \citet[][]{Comparat:2016zz} luminosity function ({\it orange dashed}). 
   We also show the results when directly using the EL-COSMOS catalog and impose the PFS magnitude and color cuts as well as the most updated flux noise as a function of wavelength ({\it blue} curves): the {\it solid (dotted)} blue curve shows the case with (without) the correction of [OII] luminosity (see Section \ref{sec:discussion}). The dip at $z\sim 1.6$ comes from features in the PFS throughput estimate, which effects are amplified in the low signal-to-noise ratio regime. 
}
\end{center}
\end{figure}

We use our catalog to estimate the expected number of galaxies targeted in PFS.
From the galaxy best-fitting template SED and HSC filter responses, we first predict $(g$, $r$, $i$, $z$, $y)$ apparent magnitudes for all galaxies in the catalog. 
Then we use the PFS selection:
\begin{equation}
    23.2<g<24.2 ~~\&~~ 0.05<g-r<0.35 ~~\&~~ {\rm SNR}_{\rm [OII]}>6.
\end{equation}
Note that this selection criterion is a slightly updated version from the original one defined in \citet{Takada:2014sf}, which was based on the J09 catalog. 
The SNR are obtained by taking the ratio between the corrected [OII] flux and estimated underlying noise. 
We correct the [OII] fluxes for residual estimation error using the results from our luminosity function analysis. We compute the average relation between the uncorrected ($L_u$) and corrected ($L_c$) luminosity by performing an abundance matching and solving for $L_c$ given $(L_u, z)$, i.e.
\begin{equation}
    \int_{L_c}^\infty \Phi_c(L,z) dL = \int_{L_u}^\infty \Phi_u(L,z) dL
    \label{eq:correctL}
\end{equation}
where $\Phi_c$ ($\Phi_u$) is the corrected (uncorrected) luminosity function. 
This assumes that the luminosity ranking is preserved. 
The noise is computed by using the PFS exposure time calculator for a $15$-minutes exposure, as described in \citet{Takada:2014sf}, but with the latest expected instrumental sensitivity.
The predicted number counts for the PFS target ELGs as a function of redshift is presented in Figure~\ref{fig:O2Nz}. 
Both corrected and uncorrected counts are shown together with the prediction from the [OII] luminosity function for a purely flux-limited selection for comparison. 
In that case, we set the flux limit to $6\overline{\sigma(F_{\lambda})}=6.3\times 10^{-17}$ erg s$^{-1}$ cm$^{-2}$, where $\overline{\sigma(F_{\lambda})}$ is the flux noise averaged over $0.6<z<2.4$ in PFS. 
The predicted flux-limited counts based on \citet{Comparat:2016zz} luminosity function are also shown in the figure. 
We find that, when averaging over $0.6<z<2.4$, our model predicts $8\%$ less galaxies than \citet{Comparat:2016zz} model.
One can see in the figure that the EL-COSMOS corrected counts for the PFS targets show a dip at $z\sim 1.6$. 
This is associated to specific features in the PFS throughput estimate which propagates into the estimated SNR. 
This effectively reduces the SNR at this redshift, mostly affecting the corrected counts that have smaller SNR on average.
By integrating the estimated PFS (corrected) counts over $0.6<z<2.4$ we find an expecting number of $3886\pm53$ galaxies per deg$^{2}$.

%~~~~~~~~~~~~~~~~~~~~~~~~~~~~~~%
%~~~~~~~~~~~~~~~~~~~~~~~~~~~~~~%

%==============================%
%==============================%
\section{Summary and Conclusion}
\label{sec:conclusion}
%==============================%
%==============================%

The majority of next-generation galaxy redshift surveys aiming at understanding the accelerated expansion of the Universe will target emission-line galaxies at intermediate redshifts, and up to about $2$. Despite this craze for emission-line galaxies, little is known about their physical properties and abundance, particularly at redshifts above unity. It is however essential to model and understand emission lines from star-forming galaxies in order to design and prepare those high-redshift galaxy redshift surveys. We try to address this question in this paper by building a value-added catalog based on the COSMOS2015 catalog of \citet{Laigle:2016fr}, which includes main emission-line fluxes in addition to the already available galaxy physical properties.  

We use an empirical but physically-motivated approach to model the galaxy emission-line fluxes in the COSMOS field, where deep multi-band photometric datasets are available. 
We perform a SED fitting to COSMOS2015 galaxies including a careful EL flux modeling based on the number of photo-ionizing photons predicted by the stellar continuum template. 
Our prediction of the intrinsic EL flux is consistent with the empirical Kennicutt relation (see Figure~\ref{fig:kennicutt}). 
We find that a simple, redshift-independent model for the dust attenuation fails to explain the observed [OII] and H$\alpha$ fluxes in zCOSMOS and 3D-HST samples at the same time. 
In addition, this failure seems correlated with inferred values of the dust attenuation for the stellar continuum. 
On the basis of these findings, we propose to introduce a redshift evolution in the dust stellar-to-nebular attenuation parameter, $f(z)=0.44+0.2z$, which is qualitatively consistent with direct observations and the semi-analytic model predictions of \citet[][]{Izquierdo-Villalba:2019aa} (Figure~\ref{fig:f_z}).
We show the performance of our EL flux modeling in section~\ref{subsec:results}. 
In general, our predictions are consistent with the observed EL fluxes in zCOSMOS and 3D-HST within a factor of two. We also discuss possible reasons which drive some discrepancies in our modeling (Figures.~\ref{fig:flux_fz} and \ref{fig:zphoto}).

Using the derived EL flux, we have been able to measure for the first time the H$\alpha$ and [OII] luminosity functions in a consistent way over a wide range of cosmic epochs, from $z=0.3$ to $z=2.5$.
We take a particular care of the impact of redshift and EL flux modeling uncertainties on the LFs. In particular, we model the uncertainties of EL luminosities in light of the reference spectroscopic samples zCOSMOS and 3D-HST (section~\ref{subsec:error}).
We model the redshift evolution of the LFs using a Schechter function with redshift-dependent parameters (section~\ref{subsec:lf_evolution}), and compare our best-fit model with previous LF measurements in the literature (section~\ref{subsec:lf_litterature}). 
We finally present our predictions of the ELG number densities in forthcoming surveys including PFS and Euclid (section~\ref{sec:discussion}).

We believe that our synthetic catalog of EL fluxes provides an informative mapping between galaxy global properties and EL fluxes that can be used to investigate galaxy star formation properties and their evolution, although there still remains limitations and uncertainties. We make the catalog as well as the modeled spectral energy distribution for all COSMOS2015 galaxies publicly available. The detailed description of the catalog release is given in Appendix~\ref{appendix:catalog}.

%==============================%
%==============================%
\section*{Acknowledgements}
%==============================%
%==============================%

We are grateful to Johan Comparat, Lucia Pozzetti, M\'ed\'eric Boquien, Yongquan Xue, and the PFS collaboration for useful discussions and Alexandre Barth\'el\'emy for preliminary work on the subject.
We also thank Yannick Roehlly for the effort to make our catalog publicly available on the database. 
Among the authors, SS, SDLT, and OI have equally and significantly contributed to this publication.
SS was supported in part by JSPS KAKENHI Grant Number JP15H05896, and by World Premier International Research Center Initiative (WPI Initiative), MEXT, Japan. 
SS was supported in part by the Munich Institute for Astro- and Particle Physics (MIAPP) which is funded by the Deutsche Forschungsgemeinschaft (DFG, German Research Foundation) under Germany's Excellence Strategy (EXC-2094-390783311).
SDLT aknowledges the  support of the OCEVU Labex (ANR-11-LABX-0060) and the A*MIDEX project (ANR-11-IDEX-0001-02) funded by the ``Investissements d'Avenir'' French government program managed by the ANR. 
OI acknowledges the funding of the ANR for the project ``SAGACE.'' 
This work is partly supported by the Centre National d'Etudes Spatiales (CNES). 
Finally, we would also like to recognize the contributions from all of the members of the COSMOS team who helped in obtaining and reducing the large amount of multi-wavelength data that are now publicly available through IRSA at http://irsa.ipac.caltech.edu/Missions/cosmos.html.
This work is based on observations taken by the 3D-HST Treasury Program (HST-GO-12177 and HST-GO-12328) with the NASA/ESA Hubble Space Telescope, which is operated by the Association of Universities for Research in Astronomy, Inc., under NASA contract NAS5-26555.
This research has made use of the ASPIC database, operated at CeSAM/LAM, Marseille, France.

%%%%%%%%%%%%%%%%%%%%%%%%%%%%%%%%%%%%%%%%%%%%%%%%%%

\bibliographystyle{mnras}
\bibliography{ms} 
%%%%%%%%%%%%%%%%%%%%%%%%%%%%%%%%%%%%%%%%%%%%%%%%%%

%%%%%%%%%%%%%%%%% APPENDICES %%%%%%%%%%%%%%%%%%%%%

\appendix
%==============================%
%==============================%
\section{Photometric redshift uncertainty}
\label{appendix:z_photo}
\begin{figure*}
\begin{center}
\includegraphics[width=0.8\columnwidth]{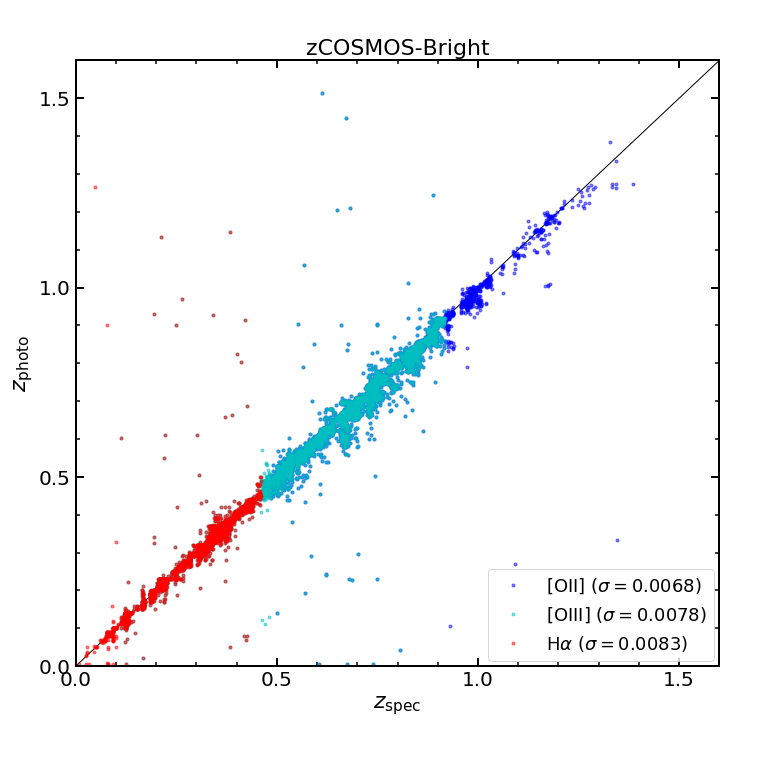}
\includegraphics[width=0.8\columnwidth]{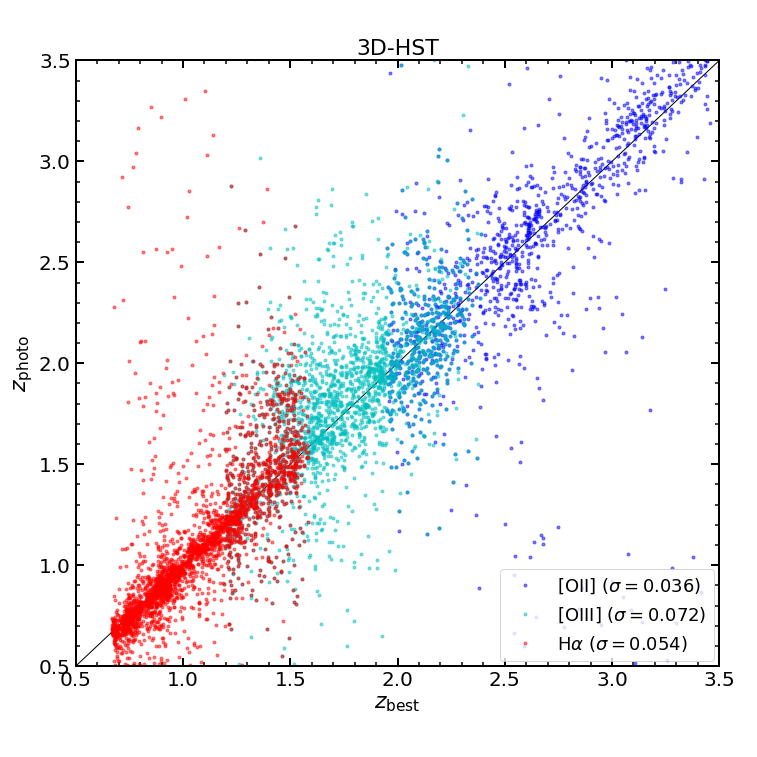}
\caption{
  \label{fig:zphoto-zspec}
  Photometric redshift in COSMOS2015 versus spectroscopic redshift in zCOSMOS (\textit{left}) and 3D-HST (\textit{right}) for our EL galaxy samples. 
  H$\alpha$, [OII], and [OIII] are shown with \textit{red}, \textit{blue}, and \textit{cyan} data points, respectively. 
  We also quote the standard deviation, $\sigma$, as $z_{\rm photo}=z_{\rm spec}\pm \sigma(1+z_{\rm spec})$. 
  }
\end{center}
\end{figure*}

%==============================%
%==============================%
Figure~\ref{fig:zphoto-zspec} shows a comparison between photometric redshift in COSMOS2015 and spectroscopically measured values. 
Here we focus on the comparison only for the EL galaxies for the zCOSMOS-Bright and 3D-HST galaxies. 
We refer readers to \citet[][]{Laigle:2016fr} for more general comparison.  
We also estimate the uncertainties of photometric redshift with the form of $z_{\rm photo}=z_{\rm spec}\pm \sigma(1+z_{\rm spec})$ where $\sigma$ we calculate $\sigma$ as the normalized median absolute deviation, $1.48\times\,{\rm median}(|z_{\rm photo}-z_{\rm spec}|/(1+z_{\rm spec}))$.
As stated in the main text, the standard deviation, $\sigma$, in 3D-HST is relatively large due to the fact that $z_{\rm best}$ does not necessarily come from the spectroscopic redshift of the emission lines.

%~~~~~~~~~~~~~~~~~~~~~~~~~~~~~~%
%~~~~~~~~~~~~~~~~~~~~~~~~~~~~~~%

%==============================%
%==============================%
\section{The impact of EL measurement uncertainties in the reference spectroscopic samples}
\label{appendix:impact_of_SN}
%==============================%
%==============================%

\begin{figure*}
\begin{center}
\hspace*{-0.7cm}
\includegraphics[width=0.9\columnwidth]{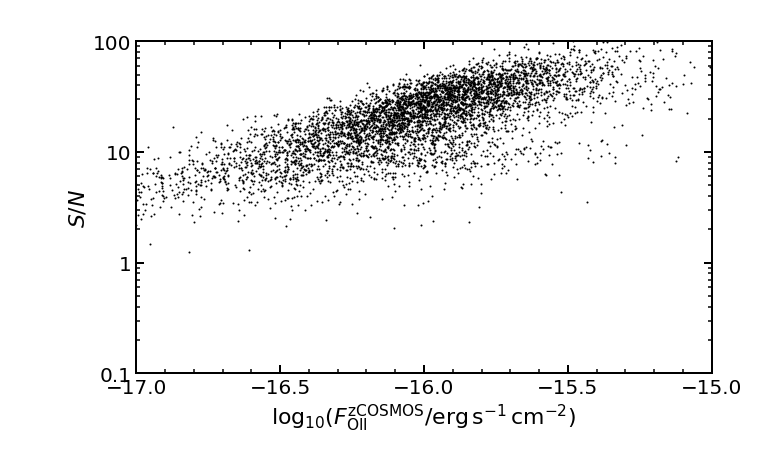}
\includegraphics[width=0.9\columnwidth]{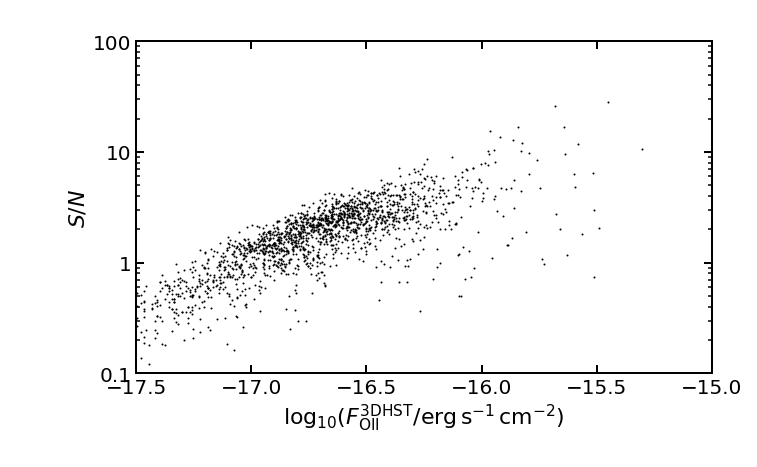}
\hspace*{-0.7cm}
\caption{
	\label{fig:SN}
    The signal-to-noise in the EL measurements for our zCOSMOS-bright (left) and 3D-HST (right) samples. 
}
\end{center}
\end{figure*}

\begin{figure*}
\begin{center}
\hspace*{-0.7cm}
\includegraphics[width=1.\textwidth]{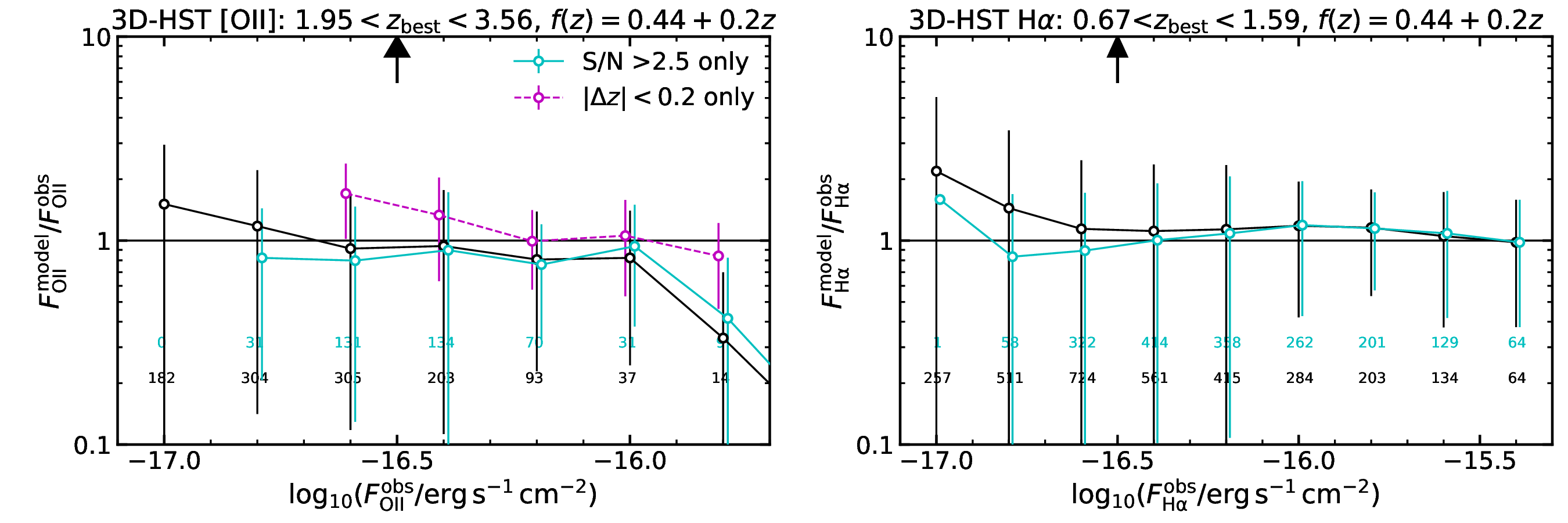}
\hspace*{-0.7cm}
\caption{
	\label{fig:SN2.5}
    The similar figure to Figure~\ref{fig:f_z} only for 3D-HST but after applying the cut with $S/N>2.5$ for the EL measurements (cyan points and curves). 
    The black solid curves are exactly the same as Figure~\ref{fig:f_z}. 
}
\end{center}
\end{figure*}

In this appendix we discuss the impact of the uncertainties on the reference spectroscopic samples. 
In Section~\ref{subsec:results}, we showed all the results without making any cuts in terms of the quality of the EL measurements in the reference spectroscopic samples, while we make a cut of $S/N>2.5$ in the luminosity function analysis. 
We have made this choice, since we are interested in the comparison of our EL model even at the very faint flux end. 
In Figure~\ref{fig:SN}, we present the signal-to-noise (S/N) of the [OII] flux measurements for zCOSMOS (left) and 3D-HST (right).
The completeness limit we defined in Figure~\ref{fig:completeness} roughly correspond to S/N of 10 and 3, respectively. 
We expect from Figure~\ref{fig:SN} that the impact of the low S/N galaxies on the flux comparison is more significant in 3D-HST than the one in zCOSMOS. 
For this reason we study its impact in the 3D-HST case in Figure~\ref{fig:SN2.5}. 
As expected, the restriction of S/N does not make a significant impact in the flux ranges beyond our completeness limit. 
The sample size of objects at the faint end is restricted due to their low S/N, which makes the statistical argument in that regime less conclusive.  

%~~~~~~~~~~~~~~~~~~~~~~~~~~~~~~%
%~~~~~~~~~~~~~~~~~~~~~~~~~~~~~~%

%~~~~~~~~~~~~~~~~~~~~~~~~~~~~~~%
%~~~~~~~~~~~~~~~~~~~~~~~~~~~~~~%

%==============================%
%==============================%
\section{EL-COSMOS catalog description}
\label{appendix:catalog}
\begin{figure*}
\begin{center}
\includegraphics[width=1\textwidth, bb=0 0 1008 756]{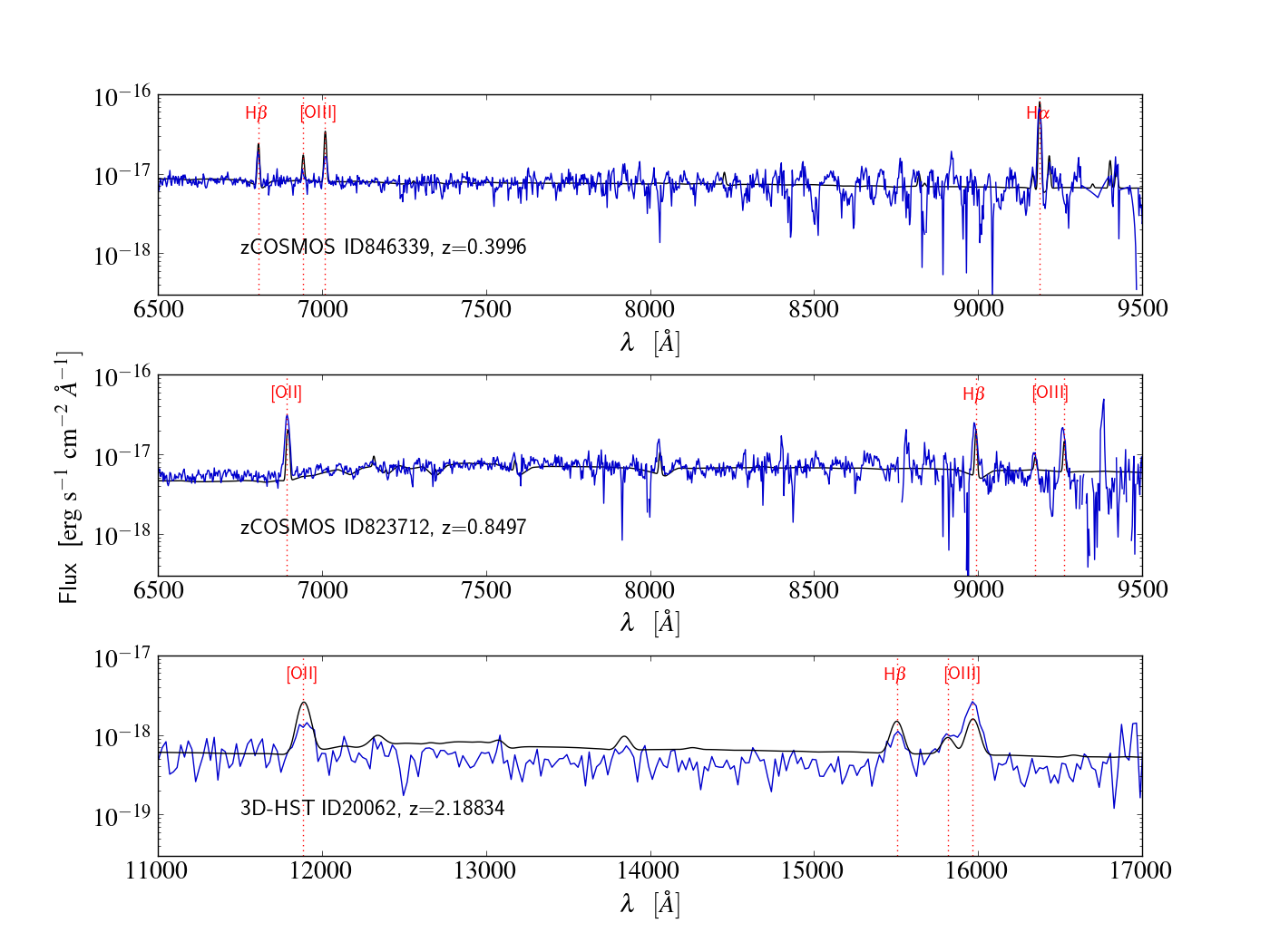}
\caption{
  \label{fig:SED}
  Three examples of our modeled spectra compared with zCOSMOS ({\it top} and {\it middle}) or 3D-HST observed spectra ({\it bottom}). 
  }
\end{center}
\end{figure*}

%==============================%
%==============================%
We make our EL-COSMOS catalog with all spectra publicly available on the ASPIC database at \url{http://cesam.lam.fr/aspic/}. 
In Table~\ref{table:public_catalog}, we summarise the available quantities in the catalog and 
refer the reader to \citet[][]{Laigle:2016fr} for the definition of the magnitude filters. 
We also show three examples of predicted spectra (black lines) in Figure~\ref{fig:SED}, which we compare to observed zCOSMOS or 3D-HST spectra (blue lines). 
Note that, for a fair comparison, we apply the aperture correction on the observed zCOSMOS spectra, and convolve our model spectra with a Gaussian filter of width 3 (35) angstroms to mimic zCOSMOS (3D-HST) spectral resolution. 
In addition, we shift the model spectra to exactly match the spectroscopic redshift. We recall that the modeled fluxes have non-negligible uncertainties and thus the catalog may be more suited for statistical studies. 

%~~~~~~~~~~~~~~~~~~~~~~~~~~~~~~%
%~~~~~~~~~~~~~~~~~~~~~~~~~~~~~~%

\begin{table*}
\begin{tabular}{|c|c|c|}
\hline
Field       & Units                                             & Description                                                 \\ \hline
ID          & -                                                 & The unique ID of a galaxy in the COSMOS2015 catalog        \\
RA          & deg                                               & Right ascension angle                                               \\
Dec        & deg                                               & Declination angle                                            \\
ZPHOT       & -                                                 & Photometric redshift from the COSMOS2015 catalog    
                     \\
MASS\_BEST  & ${\rm M}_{\odot}$                                 & Stellar mass                                               \\
SFR\_BEST   & ${\rm M}_{\odot}\,{\rm yr}^{-1}$                  & Star formation rate                                        \\
SSFR\_BEST  & ${\rm yr}^{-1}$                                   & Specific star formation rate                              \\
u\_CFHT     & AB magnitude                                     & u-band CFHT predicted magnitude                                   \\
g\_HSC      & AB magnitude                                     & g-band HSC predicted magnitude                                    \\
r\_HSC      & AB magnitude                                    & r-band HSC predicted magnitude                                    \\
i\_HSC      &  AB magnitude                                  & i-band HSC predicted magnitude                                    \\
z\_HSC      &  AB magnitude                                    & z-band HSC predicted magnitude                                    \\
y\_HSC      &  AB magnitude                                    & y-band HSC predicted magnitude                                    \\
J\_VISTA    & AB magnitude                                     & J-band VISTA predicted magnitude                                  \\
H\_VISTA    &   AB magnitude                                  & H-band VISTA predicted magnitude                                  \\
K\_VISTA    & AB magnitude                                     & K-band VISTA predicted magnitude                                  \\
{[}OII{]}   & \multicolumn{1}{c|}{${\rm erg\,s^{-1}\,{cm}^{-2}}$} & Estimated {[}OII{]} $(\lambda 3727, \lambda 3730)$ flux \\
Hb          & \multicolumn{1}{c|}{${\rm erg\,s^{-1}\,{cm}^{-2}}$} & Estimated H$\beta$ $\lambda 4863$ flux                  \\
{[}OIIIa{]} & \multicolumn{1}{c|}{${\rm erg\,s^{-1}\,{cm}^{-2}}$} & Estimated {[}OIII{]} $\lambda 4960$ flux                \\
{[}OIIIb{]} & \multicolumn{1}{c|}{${\rm erg\,s^{-1}\,{cm}^{-2}}$} & Estimated {[}OIII{]} $\lambda 5008$ flux                \\
Ha          & \multicolumn{1}{c|}{${\rm erg\,s^{-1}\,{cm}^{-2}}$} & Estimated H$\alpha$ $\lambda6565$ flux                  \\ \hline
\end{tabular}
\caption{
  \label{table:public_catalog}
  Details of data field in the released \texttt{fits} file.
  }
\end{table*}

%%%%%%%%%%%%%%%%%%%%%%%%%%%%%%%%%%%%%%%%%%%%%%%%%%
% Don't change these lines
% * <ssaito.utap@gmail.com> 2018-05-19T13:01:29.730Z:
%
% ^.
\bsp	% typesetting comment
\label{lastpage}
\end{document}